\documentclass[aps, pra, superscriptaddress, notitlepage, reprint]{revtex4-1}
\usepackage[english]{babel}
\usepackage{amsmath,amsthm}
\usepackage{amsfonts}
\usepackage[pdfborder={0 0 0}, colorlinks=true, urlcolor=blue, linkcolor=blue, citecolor=blue]{hyperref}
\usepackage{color}
\usepackage{graphicx}
\usepackage{float}
\usepackage{subfigure}
\usepackage{lipsum}

\theoremstyle{definition}

\theoremstyle{remark}

\begin{document}

\title{Manipulation and enhancement of asymmetric steering via interference effects induced by closed-loop coupling}
\author{Shasha Zheng}
\affiliation{State Key Laboratory of Mesoscopic Physics, School of Physics, Peking University, Collaborative Innovation Center of Quantum Matter, Beijing 100871, China}
\author{Fengxiao Sun}
\affiliation{State Key Laboratory of Mesoscopic Physics, School of Physics, Peking University, Collaborative Innovation Center of Quantum Matter, Beijing 100871, China}
\affiliation{Collaborative Innovation Center of Extreme Optics, Shanxi University, Taiyuan 030006, China}
\author{Yijie Lai}
\affiliation{Interdisciplinary Program of Science, National Tsing Hua University, Hsinchu, Taiwan}
\author{Qihuang Gong}
\affiliation{State Key Laboratory of Mesoscopic Physics, School of Physics, Peking University, Collaborative Innovation Center of Quantum Matter, Beijing 100871, China}
\affiliation{Collaborative Innovation Center of Extreme Optics, Shanxi University, Taiyuan 030006, China}
\author{Qiongyi He}
\email{qiongyihe@pku.edu.cn}
\affiliation{State Key Laboratory of Mesoscopic Physics, School of Physics, Peking University, Collaborative Innovation Center of Quantum Matter, Beijing 100871, China}
\affiliation{Collaborative Innovation Center of Extreme Optics, Shanxi University, Taiyuan 030006, China}

\begin{abstract}
We present a phase control method for a general three-mode system with closed-loop in coupling that drives the system into an entangled steady state and produces directional steering between two completely symmetric modes via quantum interference effects. In the scheme, two modes are coupled with each other both by a direct binary interaction and by an indirect interaction through a third intermediate damping mode, creating interference effects determined by the relative phase between the two physical interaction paths. By calculating the populations and correlations of the two modes, we show that depending on the phase, two modes can be prepared into an entangled steady state with asymmetric and directional steering even if they possess completely symmetric decoherence properties. Meanwhile, entanglement and steering can be significantly enhanced due to constructive interference, and thus more robust to thermal noises. This provides an active method to manipulate the asymmetry of steering instead of adding asymmetric losses or noises on subsystems at the cost of reducing steerability. Moreover, we show that the interference effects can also enhance and control the correlations between other pair of modes in the loop with opposite phase dependent behavior, indicating monogamy constraints for distributing correlations among multipartite. The present model could be applied in cavity optomechanical systems or in antiferromagnets where all components can mutually interact.
\end{abstract}

\maketitle

\section{Introduction}

A great deal of effort has been devoted to generate and control entanglement which is the most intrinsic feature of quantum mechanics with a variety of applications in quantum information processing~\cite{Horodecki2009}. A strict subset of entanglement called quantum steering is of particular interest~\cite{Reid2009,PhysRep2017}. The concept was originally discovered by Schr\"{o}dinger~\cite{Schrodinger1935} in response to the freaky ``spooky action-at-a-distance'' predicted by Einstein, Podolsky, and Rosen (EPR) in their famous paradox~\cite{Einstein1935,Reid1989}, describing the local measurements on one of two entangled particles can adjust (steer) the state of the other distant particle. There has been an increasing interest in EPR steering since it was rigorously defined by mathematical formulation from the perspective of quantum information~\cite{Wiseman2007,Jones2007,Cavalcanti2009}, i.e., in a network one can verify the entanglement between Alice and Bob without the requirement of trust of Bob's equipment used to perform local measurements at his node by confirming the presence of steering of Alice's system by Bob. 
This feature makes quantum steering substantial to various quantum information protocols which rely on entanglement by providing extra security~\cite{Opanchuk2014}, such as semi-sided device-independent quantum key distribution~\cite{Branciard2012,Gehring2015,Walk2016} and quantum secret sharing~\cite{Armstrong2015,Xiang2017,Kogias2017}, one-way quantum computing~\cite{Li2015}, no-cloning quantum teleportation~\cite{He2015,Reid2013,CMLi2016}, subchannel discrimination~\cite{Piani2015}, and other related protocols.

A second distinctive feature, in contrast to entanglement and Bell nonlocality, is that quantum steering implies a direction between the parties involved. The steering strengths in two directions may not be the same~\cite{Wagner2008}, especially, a special type is known as one-way EPR steering where the entangled states show steering in one direction but not in the other~\cite{Wiseman2007,Handchen2012}. The asymmetric EPR steering has attracted considerable attention recently for both theory~\cite{Olsen2008,Midgley2010,Olsen2013,Schneeloch2013,He2013,He2014,Evans2014,Bowles2014,Skrzypczyk2014,Wang2014,Reidjosab2015,QHe2015,Tan2015,Marco2015,Bowles2016,Olsen2017,Baker2018} and experiment~\cite{Handchen2012,Armstrong2015,Qin2017,Sun2016,Wollmann2016,Xiao2017,Tischler2018,Fadel2018}. One important method used in above studies to produce directional steering is making the states asymmetric by adding different amount of losses or noises on the subsystems. For example, the entangled two-qubit states which are one-way steerable have been experimentally demonstrated by passing one party into a lossy channel with a given probability of obtaining qubit~\cite{Wollmann2016,Tischler2018}. It is important to realize one-way steerable Gaussian continuous variable states referred in original EPR argument. One-way steering in the Gaussian regime (Gaussian states and Gaussian measurements) was first observed by introducing additional amounts of loss to one mode of a two-mode squeezed state (TMSS) (vacuum is coupled into the signal by a beam splitter with transmission efficiency $\eta$)~\cite{Handchen2012}, and then extended to a multipartite optical network~\cite{Armstrong2015}. The direction of Gaussian steering can be also manipulated by adding Gaussian noises to one party of TMSS and transmitting the TMSS in a lossy channel~\cite{Qin2017}. Rather than by adding asymmetric loss or noise to subsystems which in general leads to the reduction of correlation, recently, Olsen proposed a nonlinear optical apparatus to control the asymmetry of EPR steering by controlling the amplitude of signals injected into nondegenerate parametric oscillator such that the effects are intrinsic to the scheme~\cite{Olsen2017}.

In light of the exceptional importance of these developments, we provide in this paper a different mechanism to prepare and control the asymmetric steering with enhanced steerability via quantum interference effects in a general three-mode system where all three modes are mutually coupled with each other. We show that by engineering interference between two different interaction paths, i.e., a direct interaction path and an indirect interaction path induced by both being coupled to the third intermediate damping mode, two identical modes with symmetric decoherence properties can be driven into an entangled steady state with asymmetric and directional steering, meanwhile the entanglement and steerability can be effectively enhanced due to constructive interference. This provides an active method to create and control the direction of asymmetric steering without the cost of reducing steerability. Moreover, the steady-state entanglement and asymmetric steering created by the interference effects appear to be more robust against thermal noise. Finally, we have also demonstrated the opposite interference patterns of entanglement and steering between other pair of modes in the loop, indicating the monogamy constraints for distributing quantum correlations among multipartite. It is worth noticing that new quantum interference effects arising from a closed coupling loop have been recently applied to study optomechanical interferences and phonon correlations~\cite{Restrepo2017,Sun2017}, the phase effects on phonon blockade effects~\cite{Restrepo2017,Wang2017}, the transduction bandwidth of a rf-to-optical transducer ~\cite{MoaddelHaghighi2018}, and the optical nonreciprocal behavior~\cite{Metelmann2015,Xu2015,Xu2016,Xu2017} which has been experimentally implemented~\cite{Ruesink2016,Barzanjeh2017,Bernier2017}.

The remainder of this paper is organized as follows. In Sec.~\ref{Model}, we introduce a general three-mode system with closed-loop in coupling which creates interference effects determined by the relative phase of the coupling strengths between two interaction paths. In Sec.~\ref{CRITERIA}, we recall the criteria adopted in this paper to test quantum entanglement and Gaussian steering. We then investigate in Sec.~\ref{RESULTS}, the creation and control of the steady-state entanglement and quantum steering by the interfering channels and compare its performance with the results achieved by applying only one interaction path. The results show the enhancement of correlations, the effective control of the asymmetric steering, and the robustness to thermal noises via interference effects. Finally, we summarize our results in Sec.~\ref{SUMMARY}.

\section{Model}\label{Model}
\begin{figure}[htbp!]
\centering
\includegraphics[width=0.5\columnwidth]{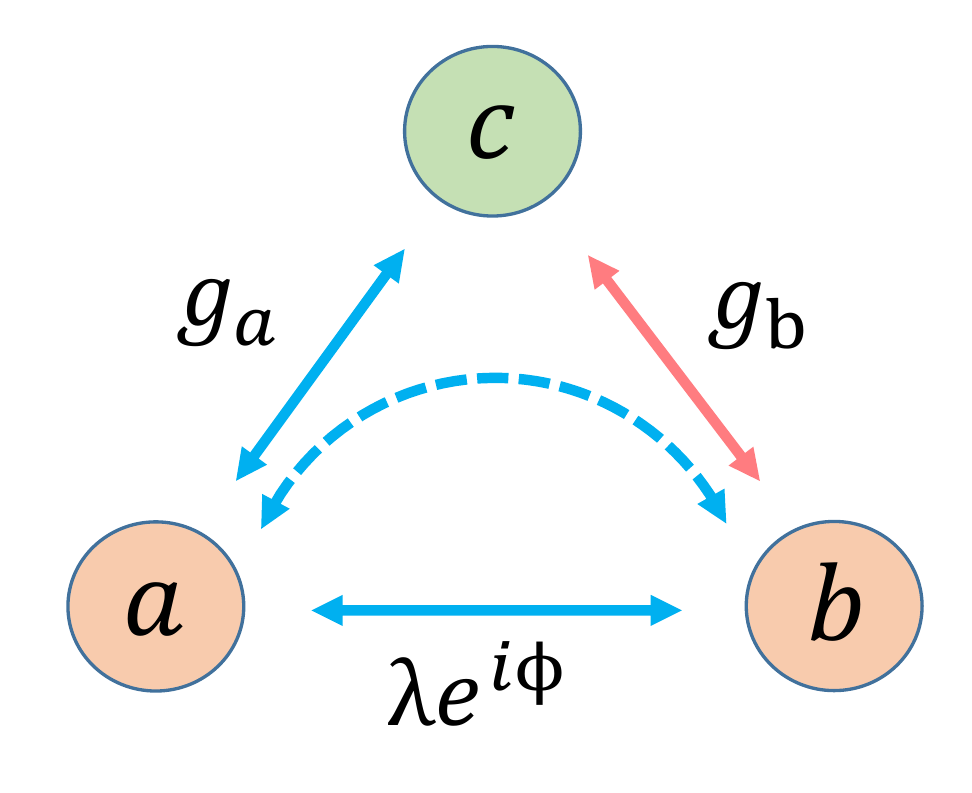}
\caption{Schematic diagram of a general three-mode system with closed coupling loop. Modes $a$ and $b$ interact directly via a two-mode squeezing interaction with strength $\lambda$; at the same time, they are coupled to an intermediate damped mode $c$ resulting in an indirect interaction path between them indicated by the blue dashed curve. The relative phase $\phi$ between two paths can be engineered to control and improve the correlation between modes $a$ and $b$. }
\label{scheme}
\end{figure}
We consider a three-mode system represented by the annihilation operators $a$, $b$, $c$, in which modes $a$ and $b$ interact with each other in two distinct paths, i.e., the direct coupling path resulting from a two-mode squeezing interaction and the induced indirect interaction path obtained by simply coupling to the third intermediate damping mode $c$. The interaction of the system is thus described by a general form of Hamiltonian (with $\hbar=1$)
\begin{eqnarray}
H=&&\lambda e^{i\phi} a^\dagger b^\dagger+\lambda e^{-i\phi}ab + g_a(e^{i\phi_a}c^\dagger a^\dagger+e^{-i\phi_a}ca)\nonumber \\
&&+g_b(e^{i\phi_b}c^\dagger b+e^{-i\phi_b}cb^\dagger),
\end{eqnarray}
where $\lambda,\ g_a,\ g_b$ are the effective coupling strengths with individual phase $\phi,\ \phi_a,\ \phi_b$ controlled by the driving laser fields. This creates a three-mode closed-loop in the coupling, as illustrated in Fig.~\ref{scheme}, which will give rise to interference effects and phase dependence of the dynamics of the modes. Since the interference is dependent on the relative phase between the two interaction paths, without loss of generality, we can absorb the phases $\phi_a$ and $\phi_b$ into $\phi$ by redefining the operators $a$ and $b$. In the following we will assume $\phi_a=\phi_b=0$ and treat $\phi$ as the relative phase in the interference for simplicity. This coupling loop could be realized in three-mode optomechnical systems~\cite{Sun2017}, or in antiferromagnets with magnon-photon coupling~\cite{Yuan2017}.

Substituting the Hamiltonian to the Heisenberg equation and taking into account the dissipation-fluctuation processes, we get the quantum Langevin equations (QLEs) for each mode
\begin{eqnarray}\label{Langevin}
\dot{a}&=&-\kappa_aa-i\lambda e^{i\phi}b^\dagger-ig_ac^\dagger-\sqrt{2\kappa_a}a^{in},\nonumber \\
\dot{b}&=&-\kappa_bb-i\lambda e^{i\phi} a^\dagger-ig_bc-\sqrt{2\kappa_b}b^{in},\nonumber \\
\dot{c}&=&-\gamma_cc-i(g_bb+g_aa^\dagger)-\sqrt{2\gamma_c}c^{in},
\end{eqnarray}
where $\kappa_a, \ \kappa_b,\ \gamma_c$ are the damping rates of modes $a$, $b$, $c$, respectively, $a^{in}$, $b^{in}$ and $c^{in}$ are the input quantum noises with zero average value which are taken to be statistically independent with nonzero $\delta$ correlated functions, $
\langle a^{in}\left(t\right)a^{in\dagger}\left(t'\right)\rangle=\delta(t-t')$, $\langle b^{in}\left(t\right)b^{in\dagger}\left(t'\right)\rangle =\delta(t-t')$, $\left<c^{in}\left(t\right)c^{in\dagger}\left(t'\right)\right> = (\bar{n}_{th}+1)\delta(t-t')$, $\langle c^{in\dagger}\left(t\right)c^{in}\left(t'\right)\rangle = \bar{n}_{th}\delta(t-t')$, where
$\bar{n}_{th}=[\exp\frac{\hbar\omega_c}{k_BT}-1]^{-1}$ is mean thermal occupation number of the the intermediate mode $c$ at the frequency $\omega_c$, $T$ is the temperature of the surrounding environment, and $k_B$ is the Boltzmann constant. Here we assume that modes $a$ and $b$ are in the ordinary zero-temperature environment ($\bar{n}_{th_a}=\bar{n}_{th_b}=0$), whereas the intermediate mode is in a thermal state with nonzero temperature. We will also show the influence of noises when all three modes are thermally excited.

We are interested in the phase dependent effects on the generation and control of Gaussian entanglement and steering between modes $a$ and $b$, so define the quadrature operators of modes $X_j=(j+j^\dagger)/\sqrt{2}$, $Y_j=(j-j^\dagger)/\sqrt{2}i$, and corresponding Langevin noise operators $X_j^{in}=(j^{in}+j^{in\dagger})/\sqrt{2}$, $Y_j^{in}=(j^{in}-j^{in\dagger})/\sqrt{2}i$, $(j=a,b,c)$. The QLEs (\ref{Langevin}) can be thus written as
\begin{equation}\label{Langevin_quadrature}
\dot{u}=Mu-\Lambda u^{in},
\end{equation}
where $u=(X_{a},Y_{a},X_{b},Y_{b},X_{c},Y_{c})^T$ (the superscript $T$ denotes the transposition), the diagonal damping matrix $\Lambda=\text{diag}(\sqrt{2\kappa_a},\sqrt{2\kappa_a},\sqrt{2\kappa_b},\sqrt{2\kappa_b},\sqrt{2\gamma_c},\sqrt{2\gamma_c})$, and the corresponding vector of noises $u^{in}=(X_{a}^{in},Y_{a}^{in},X_{b}^{in},Y_{b}^{in},X_{c}^{in},Y_{c}^{in})^T$. The $6\times6$ drift matrix $M$ is the coefficient matrix of the system, which reads
 \begin{equation*}
M=-\left(
   \begin{array}{cccccc}
     \kappa_a & 0 & -\lambda\sin{\phi}& \lambda\cos{\phi} & 0 & g_a\\
     0 & \kappa_a & \lambda\cos{\phi}& \lambda\sin{\phi} & g_a & 0 \\
     -\lambda\sin{\phi} & \lambda\cos{\phi} & \kappa_b & 0 & 0 & -g_b \\
     \lambda\cos{\phi} & \lambda\sin{\phi} & 0 & \kappa_b & g_b & 0\\
     0 & g_a & 0 & -g_b & \gamma_c & 0\\
     g_a & 0 & g_b & 0 & 0 & \gamma_c
   \end{array}
 \right).
 \end{equation*}
The system is stable only when the real parts of all the eigenvalues of matrix $M$ are negative, and the stability condition can be derived from Routh-Hurwitz criterion \cite{Dejesus1987}. The general stability condition depending on the relative phase is too complex, which can be simply given for the cases of $\phi=\frac{\pi}{2}+n\pi$ $(n \in \mathbb{Z})$,
 \begin{eqnarray}
&&\kappa_a\kappa_b\gamma_c-g_a^2\kappa_b+g_b^2\kappa_a-\lambda^2\gamma_c >0, \nonumber \\
&&(\kappa_a+\kappa_b)(\kappa_a+\gamma_c)(\kappa_b+\gamma_c)\nonumber\\
&&- g_a^2(\kappa_a+\gamma_c)+g_b^2(\kappa_b+\gamma_c)-\lambda^2(\kappa_a+\kappa_b)>0.
 \label{stability condition}
 \end{eqnarray}

\section{CRITERIA FOR ENTANGLEMENT AND EPR STEERING}\label{CRITERIA}
 
The steady state of the system is a zero-mean Gaussian state because of the Gaussian nature of the input quantum noises $a^{in}$, $b^{in}$ and $c^{in}$ and the linearized dynamics, which can be fully characterized by its covariance matrix (CM) $V$ with components $V_{lm}=[\langle u_l(\infty)u_m(\infty)+u_m(\infty)u_l(\infty)\rangle]/2$ $(l,m=1,2,\cdots,6)$, where $u_l(t)$ is the $l$th quadrature component of $u(t)$.
The steady-state CM can be determined by solving the Lyapunov equation $
 MV+V M^{T}=-D$, where the diffusion matrix $D$ characterizes the stationary noise correlations and is defined through $\langle v_l(t)v_m(t')+v_m(t')v_l(t)\rangle/2=D_{lm}\delta(t-t')$ ($v=-\Lambda u^{in}$)~\cite{Genes2008,Barzanjeh2011}, such that here $D=\text{diag}(\kappa_a,\kappa_a,\kappa_b,\kappa_b,(2\bar{n}_{th}+1)\gamma_c,(2\bar{n}_{th}+1)\gamma_c)$. The Lyapunov equation is a linear equation for $V$ and can be straightforwardly solved. 
  
Since we are specifically interested in the entanglement and EPR steering created between modes $a$ and $b$, it is enough to consider the reduced CM
\begin{equation}
V=\left(
   \begin{array}{cc}
   V_a & V_{ab}\\
   V_{ab}^T & V_b
   \end{array}
   \right), 
\end{equation}
where the submatrices $V_a$ and $V_b$ are corresponding to the reduced states of modes $a$ and $b$, respectively. In order to measure the entanglement, we adopt the logarithmic negativity $E_N=\max{\{0,-\ln{2\eta^{-}}\}}$ \cite{Vidal2002,Adesso2004}, where $\eta^{-}\equiv\sqrt{\Sigma(V)-[\Sigma(V)^2-4\det{V}]^{1/2}}/\sqrt{2}$, with $\Sigma(V)\equiv\det{V_a}+\det{V_b}-2\det{V_{ab}}$. Therefore, a Gaussian state is entangled if and only if $\eta^{-}<1/2$, and larger value of $E_N$ implies stronger entanglement between modes.

To signify EPR steering, we adopt the computable measure of Gaussian steering proposed in Ref.~\cite{Kogias2015} for arbitrary bipartite Gaussian states under Gaussian measurements. The steering in the direction from mode $a$ to mode $b$ ($\mathcal{G}^{a \rightarrow b}$) and in the opposite direction ($\mathcal{G}^{b\rightarrow a}$) are quantified by
 \begin{eqnarray}
&& \mathcal{G}^{a\rightarrow b}=\max{\{0,S(2V_a)-S(2V)\}}, \nonumber \\
&& \mathcal{G}^{b\rightarrow a}=\max{\{0,S(2V_b)-S(2V)\}},
 \end{eqnarray}
 where $S(\sigma)=\frac{1}{2}\ln{\det{\sigma}}$ is the R\'enyi-2 entropy. $\mathcal{G}^{a \rightarrow b}>0$ ($\mathcal{G}^{b \rightarrow a}>0$) implies that mode $a$ ($b$) can steer mode $b$ ($a$) by Gaussian measurements, and its value quantifies the degree of steering, i.e., the higher the value of $\mathcal{G}$ becomes, the stronger the Gaussian steerability appears.
 
For the present system, the steady-state values $\langle a^2\rangle=\langle b^2\rangle=\langle a^\dagger b\rangle=0$, such that the conditions to satisfy $E_N>0$, $\mathcal{G}^{a\rightarrow b}>0$ and $\mathcal{G}^{b\rightarrow a}>0$ can be also expressed in terms of correlation-based inequalities, respectively~\cite{Hillery2006,Cavalcanti2011}, 
 \begin{eqnarray}
 |\langle ab\rangle| &>&\sqrt{\langle a^\dagger a\rangle\langle b^\dagger b\rangle}, \nonumber\\
 |\langle ab\rangle|&>&\sqrt{\langle b^\dagger b\rangle(\langle a^\dagger a\rangle+1/2)}, \label{G12}\nonumber\\
 |\langle ab\rangle|&>&\sqrt{\langle a^\dagger a\rangle(\langle b^\dagger b\rangle+1/2)}. \label{G21}
 \end{eqnarray}
The steady-state solution of populations $\langle a^\dagger a\rangle, \ \langle b^\dagger b\rangle $ and correlation $\langle ab\rangle$ at $\phi=\frac{\pi}{2}+n\pi\ (n \in \mathbb{Z})$ is detailed in Appendix~\ref{solution}, and the derivation of the inequalities~(\ref{G21}) is given in Appendix~\ref{derivation}.

\section{Control and enhancement of quantum correlations via interference}\label{RESULTS}
\subsection{Phase-sensitive quantum entanglement and directional steering}

Both the direct two-mode squeezing interaction and the induced indirect interaction by being coupled to the third intermediate mode with beam-splitter-type coupling and parametric-down-conversion-type coupling, respectively, can create entanglement between modes $a$ and $b$. Thus the superposition of these two physical interaction paths may create quantum interference effects which depend on the relative phase of the coupling strengths. In the following we will investigate the phase dependent effects on the creation of quantum entanglement and EPR steering. 
\begin{figure}
\includegraphics[width=0.75\columnwidth]{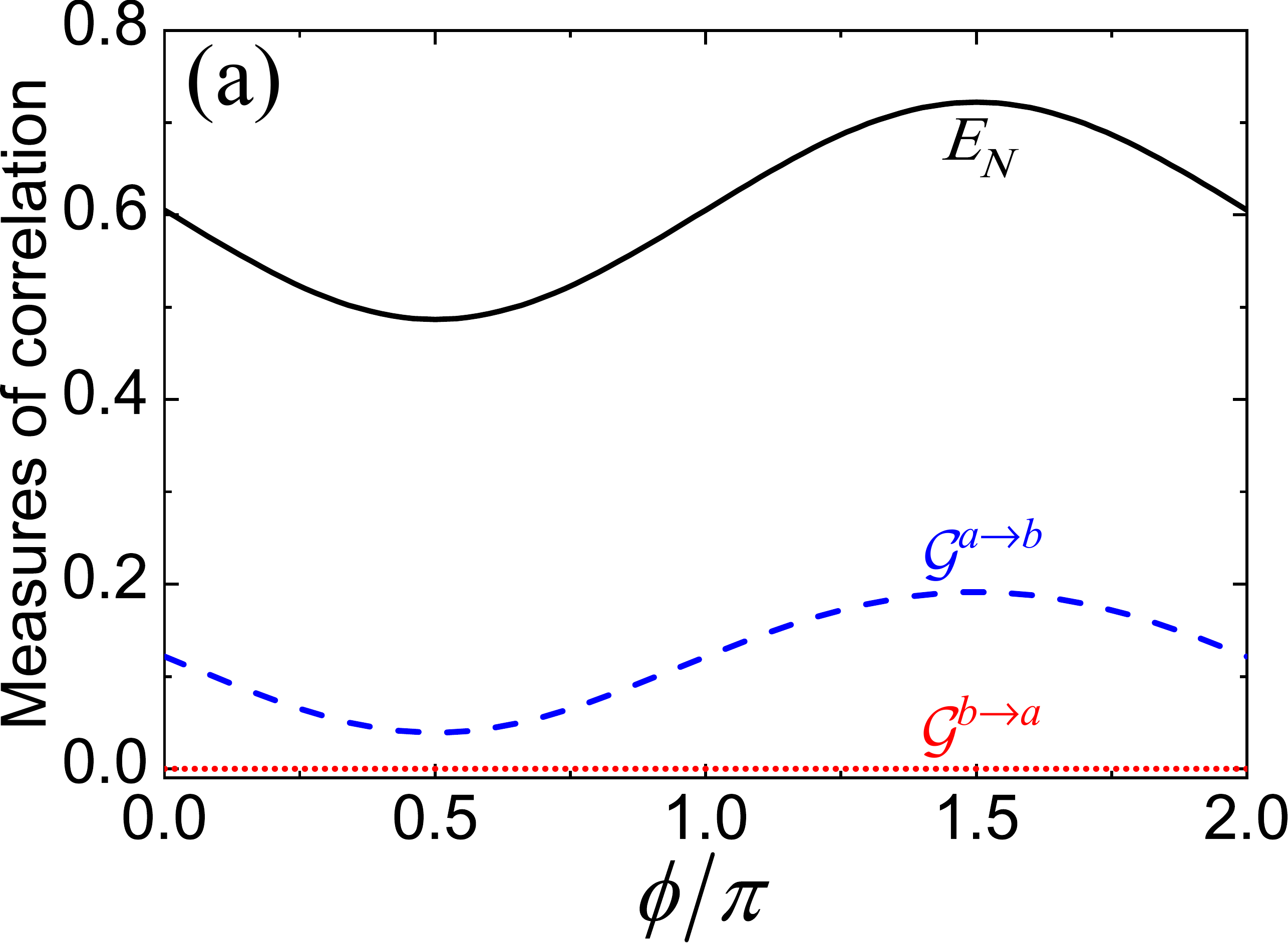}
\includegraphics[width=0.75\columnwidth]{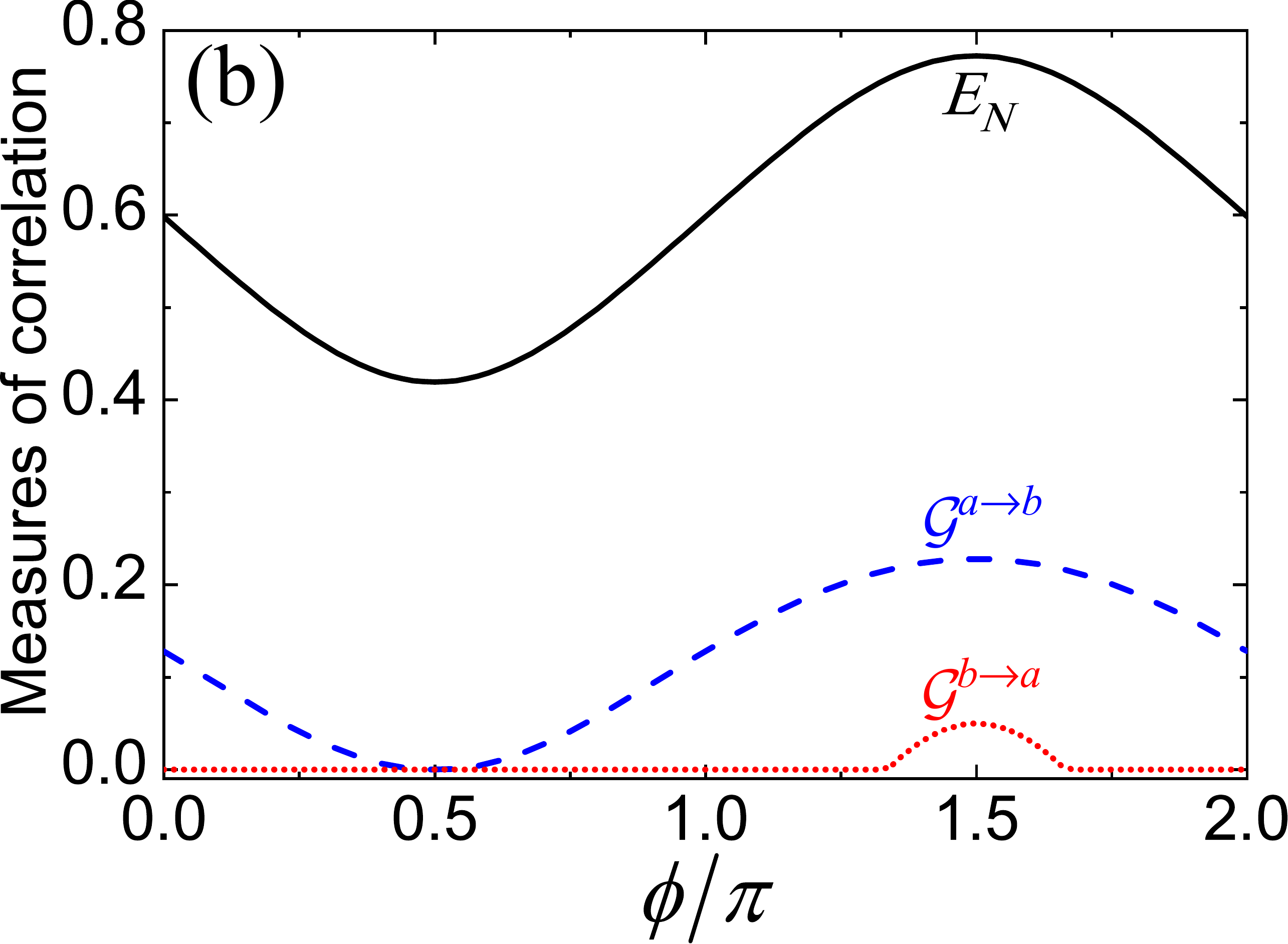}
\caption{Stationary entanglement ($E_N$, black solid) and quantum steering between two modes $a$ and $b$ ($\mathcal{G}^{a\rightarrow b}$: blue dashed, $\mathcal{G}^{b\rightarrow a}$: red dotted) as a function of the relative phase $\phi$ of interfering channels, when (a) $\lambda=0.4\kappa$ and (b) $\lambda=0.605\kappa$. Other parameters are $\kappa_a=\kappa_b=\kappa,\ g_a=3.2\kappa,\ g_b=5\kappa,\gamma_c=2\kappa,\ \bar{n}_{th}=0$.}
\label{bbr_phase}
\end{figure}

The measures of entanglement ($E_N$) and EPR steering ($\mathcal{G}$) vary with $\phi$, as depicted in Fig.~\ref{bbr_phase}. Apparently, entanglement evolves with $\phi$ periodically (black solid curve), maximized when phase $\phi=(2n+3/2)\pi$ ($n \in \mathbb{Z}$) by constructive interference and minimized at $\phi=(2n+1/2)\pi$ by destructive interference. Interestingly, Gaussian steering in two directions behave asymmetrically with phase, that is, steering from mode $a$ to mode $b$ quantified by the parameter $\mathcal{G}^{a\rightarrow b}$ (blue dashed curve) exhibits the same phase-depdent behavior as that of the entanglement measure $E_N$, whereas steering in the other direction measured by the parameter $\mathcal{G}^{b\rightarrow a}$ (red dotted curve) behaves in a completely different way. Specifically, steering in this direction doesn't exist for the case of $\lambda=0.4\kappa$ shown in Fig.~\ref{bbr_phase}(a), such that one-way steering in the direction of $a\rightarrow b$ occurs over the whole period of $\phi$. For larger value of $\lambda=0.605\kappa$, $\mathcal{G}^{b\rightarrow a}$ reaches a maximal value at $\phi=(2n+3/2)\pi$ as well, and we can get rich properties of Gaussian steering, e.g., the overall state's asymmetry is stepwise driven through the one-way regime ($\mathcal{G}^{a\rightarrow b}>0$ but $\mathcal{G}^{b\rightarrow a}=0$), no-way regime ($\mathcal{G}^{a\rightarrow b}=0$ and $\mathcal{G}^{b\rightarrow a}=0$), and two-way regime ($\mathcal{G}^{a\rightarrow b}>0$ and $\mathcal{G}^{b\rightarrow a}>0$), and finally one-way regime again over one period of phase, as shown in Fig.~\ref{bbr_phase}(b). 

For the parameters studied here, we can find that the steerability from $a$ to $b$ is always larger than the steerability from $b$ to $a$. This can be understood from the inequalities (\ref{G21}) and the solutions given in Eq.~\ref{correlation_term}. When the two modes have same dissipation rates $\kappa_a=\kappa_b=\kappa$ and mode $c$ is at zero temperature $\bar{n}_{th}=0$, from Eq.~\ref{correlation_term} we can see that the photon fluctuation of mode $a$ is always larger than that in mode $b$, i.e., $\langle a^\dagger a\rangle>\langle b^\dagger b\rangle$, and therefore the condition given in Eq.~(\ref{G12}) to confirm steering $a\rightarrow b$ is easier to satisfy than that required for demonstrating steering $b\rightarrow a$.

Now we come to the first important result of this paper. Two identical modes which have completely symmetric decoherence $\kappa_a=\kappa_b=\kappa$ and $\bar n_{th,a}=\bar n_{th,b}=0$ can be still prepared into an entangled state with asymmetric and directional steering by adjusting the phase of the interfering channels. This reveals the inherent asymmetric nature of quantum steering with respect to the two parties involved. Comparing with the way to produce one-way steering in earlier studies by adding losses or thermal noises on one subsystem~\cite{Handchen2012,Armstrong2015,Qin2017} with the cost of reducing correlation, engineering the interference is an active method which creates asymmetric steering and at the same time enhances the steerability.

\subsection{Performance comparison with correlations created by only one interaction path}
\begin{figure}
\includegraphics[width=1.0\columnwidth]{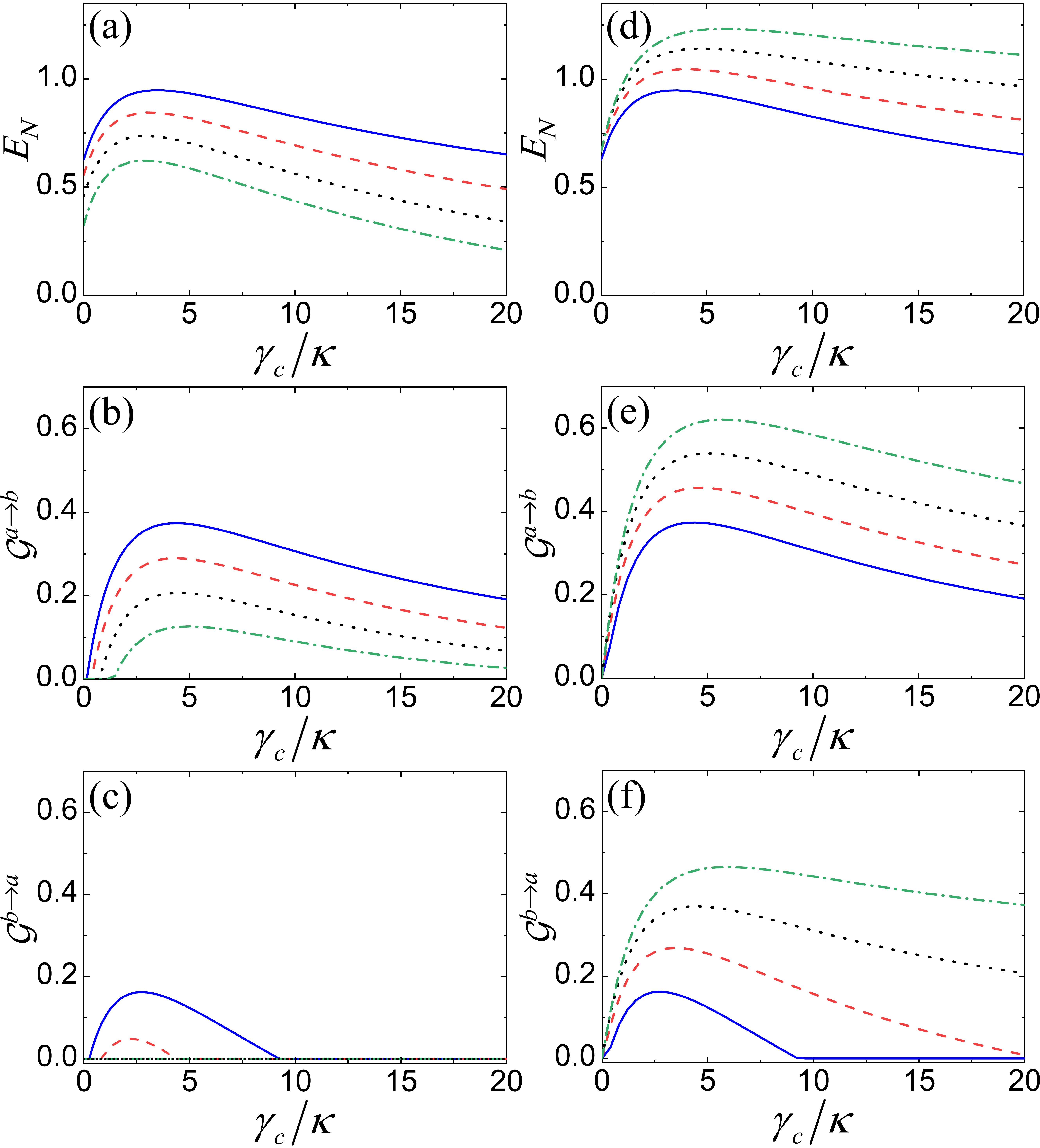}
\caption{The measures of entanglement ($E_N$) and steering ($\mathcal{G}^{a\rightarrow b}$, $\mathcal{G}^{b\rightarrow a}$) between modes $a$ and $b$ versus the damping rate $\gamma_c/\kappa$ for the cases of creating correlations only by the indirect interaction path $\lambda=0$ (blue solid), and by interfering channels when $\lambda=0.5\kappa$ (red dashed), $\lambda=\kappa$ (black dotted), $\lambda=1.5\kappa$ (green dash-dotted). The destructive and constructive interference at phase $\phi=\pi/2$ and $\phi=3\pi/2$ lead to the reduction (a-c) and enhancement (d-f) of correlations, respectively. Other parameters are $g_a=8.3\kappa,\ g_b=10\kappa,\ \bar{n}_{th}=0$.}
\label{bbr_dissipation_gamma}
\end{figure}

To clearly indicate the performance of control and enhancement of entanglement and steering via interfering channels, we compare with the results achieved via only one of two interaction paths. 

First, the numerical results of the dependence of entanglement and steering on the damping rate of the third intermediate mode $\gamma_c/\kappa$ are given in Fig.~\ref{bbr_dissipation_gamma} for different coupling strengths of the direct interaction path $\lambda$. Blue solid curves present the results achieved only by the indirect coupling path, i.e., $\lambda=0$, other curves show the correlations achieved by the interfering channels with phase $\phi=\pi/2$ [Figs.~\ref{bbr_dissipation_gamma}(a-c)] and with phase $\phi=3\pi/2$ [Figs.~\ref{bbr_dissipation_gamma}(d-f)] for $\lambda=0.5\kappa$ (red dashed), $\kappa$ (black dotted), $1.5\kappa$ (green dash-dotted), respectively. It can be seen that the correlations are reduced due to destructive interference of two interaction paths at $\phi=\pi/2$. Comparing the steering in two directions shown in Figs.~\ref{bbr_dissipation_gamma}(b) and (c), increasing value of $\lambda$ leads to the stronger destructive interference effects on reducing the correlation between two modes, but it features one-way steering in a wider parameter range. For instance, when $\lambda=\kappa$ shown by the black dotted lines, the steering becomes possible only in one direction $a\rightarrow b$ for almost all values of $\gamma_c$, while for not accounting for the direct interaction path ($\lambda=0$, blue solid), one-way steering is only possible for strong mechanical damping rate $\gamma_c\gg\kappa$. 

\begin{figure}
\includegraphics[width=0.8\columnwidth]{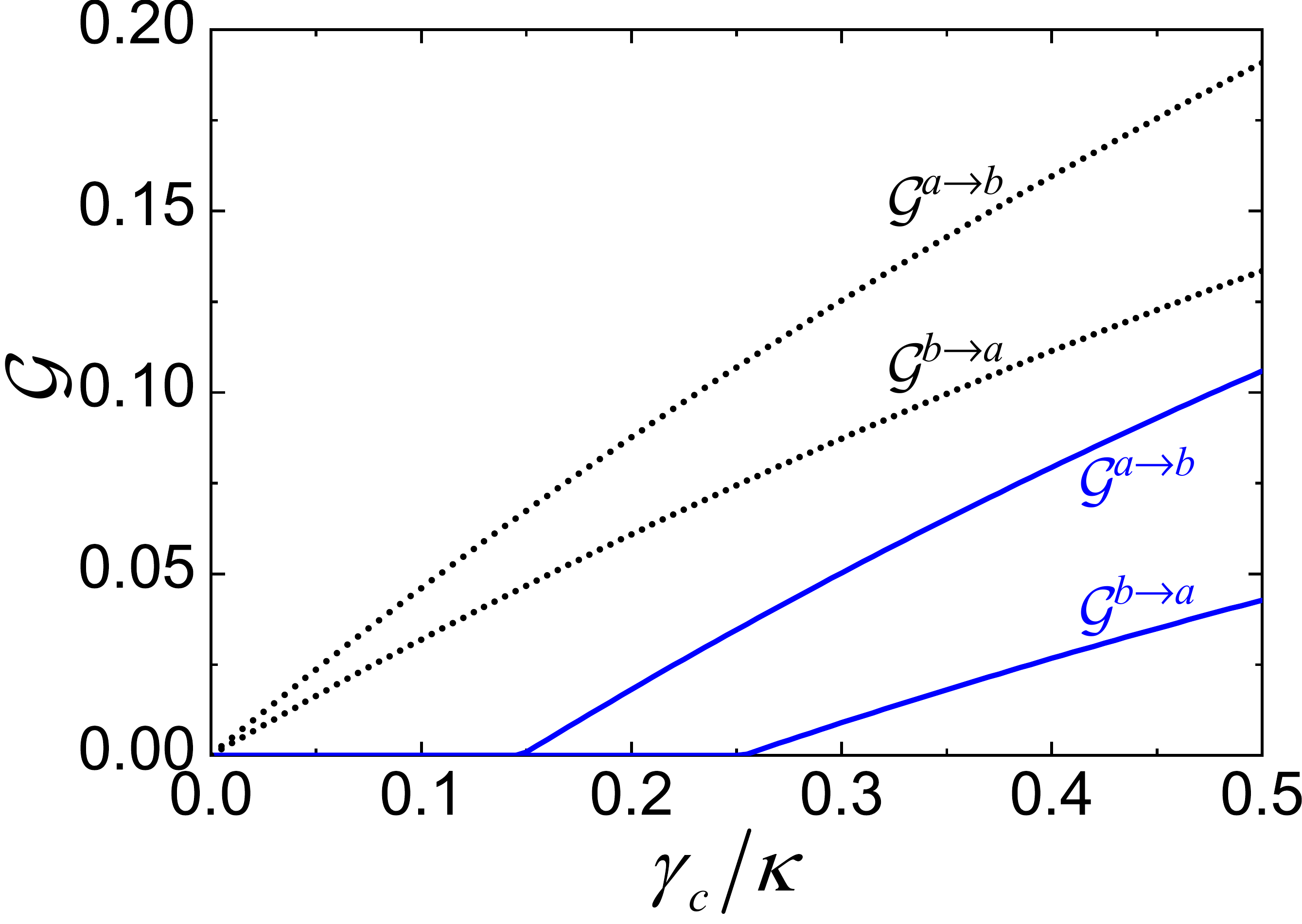}
\caption{Comparison of Gaussian steering in the regime of $\gamma_c\ll\kappa$ achieved by the indirect path only ($\lambda=0$, blue solid) and by the interfering channels with constructive interference ($\lambda=\kappa,\ \phi=3\pi/2$, black dotted). Other parameters are used as those in Fig.~\ref{bbr_dissipation_gamma}.}
\label{comparision}
\end{figure}

When the interfering channels with phase $\phi=3\pi/2$, as shown in Figs.~\ref{bbr_dissipation_gamma}(d-f), the degree of entanglement and steering are remarkably enhanced due to the constructive interference of two interaction paths comparing with the result achieved only by the indirect interaction path ($\lambda=0$, blue solid). For this case, we can get stronger two-way steering over a wider parameter range, which has been proven to be a necessary resource required for teleporting a coherent state with fidelity beyond the no-cloning threshold~\cite{He2015}. The different degree of steerability in two directions also provides the asymmetric guaranteed key rate achievable within a practical one-sided device-independent QKD~\cite{Kogias2015}. In the regime of $\gamma_c\gg\kappa$, the steering in the direction $\mathcal{G}^{b\rightarrow a}$ becomes lost. Thus, the constructive interference can also manipulate the direction of steering from asymmetric two-way to one-way. The steady-state steering generated only by the indirect interaction path has been studied in Ref.~\cite{Tan2015}, where the authors concluded that in the regime $\gamma_c\ll\kappa$ the steady entangled states are definitely not steerable in any direction when two optical modes have same damping rates $\kappa_a=\kappa_b=\kappa$, while only the mode with larger dissipation rate can be steered by the other one. Our result, however, shows that by applying the interfering channels the steering in both directions may be achieved in this regime due to the constructive interference, as indicated by the black dotted curves in Fig.~\ref{comparision}. 

Second, we compare with the results obtained when only the direct interaction path exists. In this case, the photon fluctuation and correlation expressions given by Eq.~(\ref{correlation_term}) in Appendix can be simplified as
\begin{eqnarray}
\langle a^\dagger a\rangle&=&\frac{\kappa_b\lambda^2}{(\kappa_a+\kappa_b)(\kappa_a\kappa_b-\lambda^2)}, \nonumber \\
\langle b^\dagger b\rangle&=&\frac{\kappa_a\lambda^2}{(\kappa_a+\kappa_b)(\kappa_a\kappa_b-\lambda^2)}, \nonumber \\
\langle ab\rangle&=&\frac{\kappa_a\kappa_b\lambda(\sin{\phi}-i\cos{\phi})}{(\kappa_a+\kappa_b)(\kappa_a\kappa_b-\lambda^2)}.
\end{eqnarray}
The conditions to confirm the presence of steering given in Eq.~(\ref{G12}) then reduce to
\begin{eqnarray}\label{G_coherent}
&&\mathcal{G}^{a\rightarrow b}>0:(\kappa_b-\kappa_a)(\kappa_a\kappa_b-\lambda^2)>0, \nonumber \\
&&\mathcal{G}^{b\rightarrow a}>0:(\kappa_a-\kappa_b)(\kappa_a\kappa_b-\lambda^2)>0.
\end{eqnarray}
We can see from Eq. (\ref{G_coherent}) that it is not possible to create steering in both directions when two modes possess symmetric dissipation rate, $\kappa_a=\kappa_b$. Since the stability condition of the system for this case requires $\lambda^2<\kappa_a\kappa_b$, then $\mathcal{G}^{a\rightarrow b}>0$ when $\kappa_b>\kappa_a$, and $\mathcal{G}^{b\rightarrow a}>0$ when $\kappa_a>\kappa_b$. This means that one can only produce one-way steering from the mode with lower damping rate to the mode with higher decay rate. This observation consists with the conclusion made in earlier studies~\cite{Tan2015,Handchen2012,Reidjosab2015} that the decoherence with no thermal noise on the steering system has substantial effect more than that on the steered system, such that the condition $\kappa_b>\kappa_a$ may destroy (preserve) steering in the direction $b\rightarrow a$ ($a\rightarrow b$) where mode $b$ possessing higher decoherence acts as steering (steered) party, respectively. 
\begin{figure}
\centering
\includegraphics[width=0.8\columnwidth]{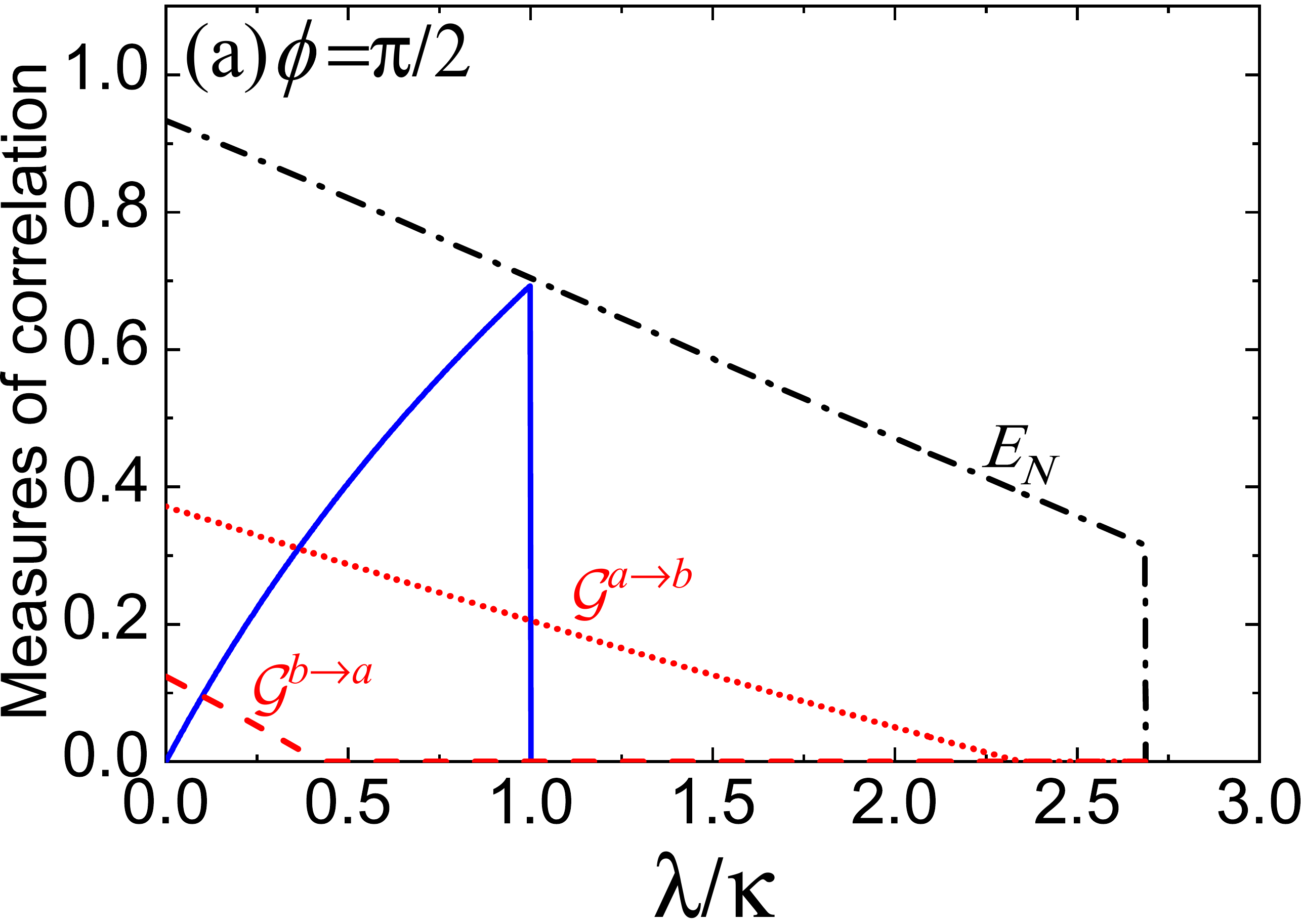}
\includegraphics[width=0.8\columnwidth]{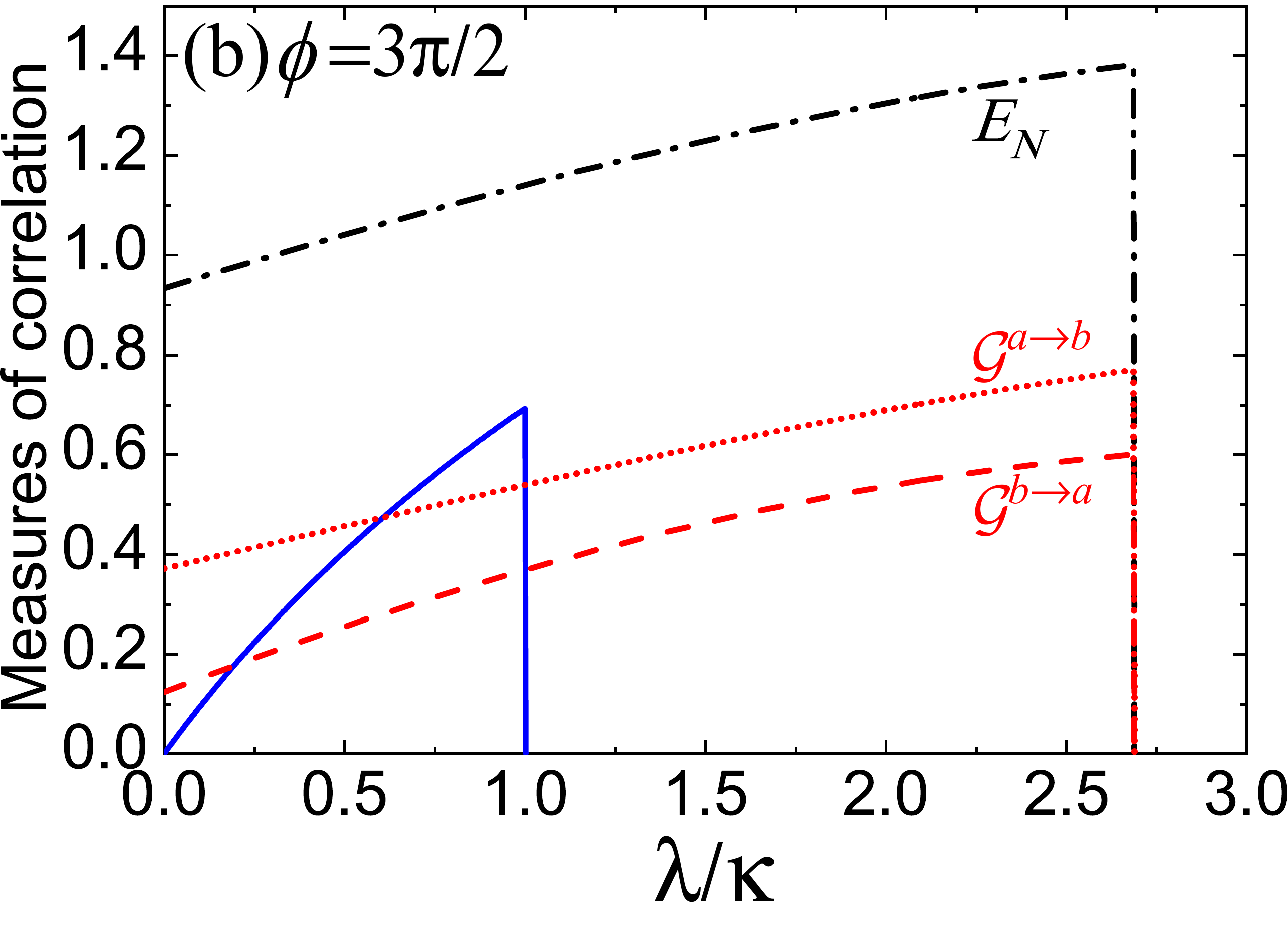}
\caption{The comparison of entanglement and steering achieved by only one direct interaction path (blue solid) and by the interfering channels at phase (a) $\phi=\pi/2$ and (b) $\phi=3\pi/2$, respectively. The entanglement $E_N$ (black dash-dotted) and steering in two directions $\mathcal{G}^{a\rightarrow b}$ (red dotted), $\mathcal{G}^{b\rightarrow a}$ (red dashed) are notably enhanced, meanwhile, the range of steady-state solution is also remarkably broadened when two interaction paths superpose and interference happens. Other parameters are $g_a=8.3\kappa,\ g_b=10\kappa,\gamma_c=5\kappa, \bar{n}_{th}=0$.}
\label{bbr_coherent_lambda0}
\end{figure}
\begin{figure}
\centering
\includegraphics[width=0.8\columnwidth]{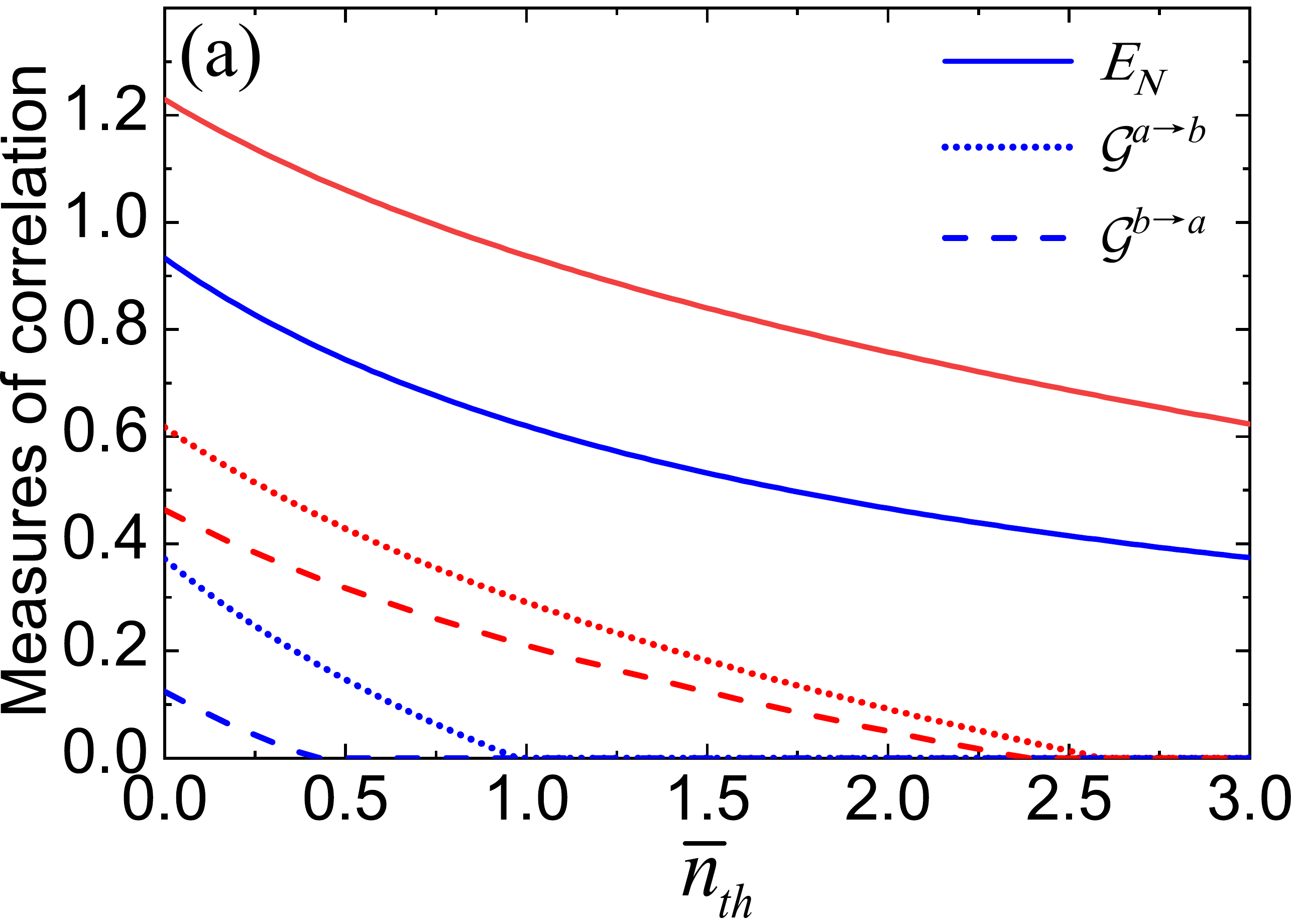}
\includegraphics[width=0.8\columnwidth]{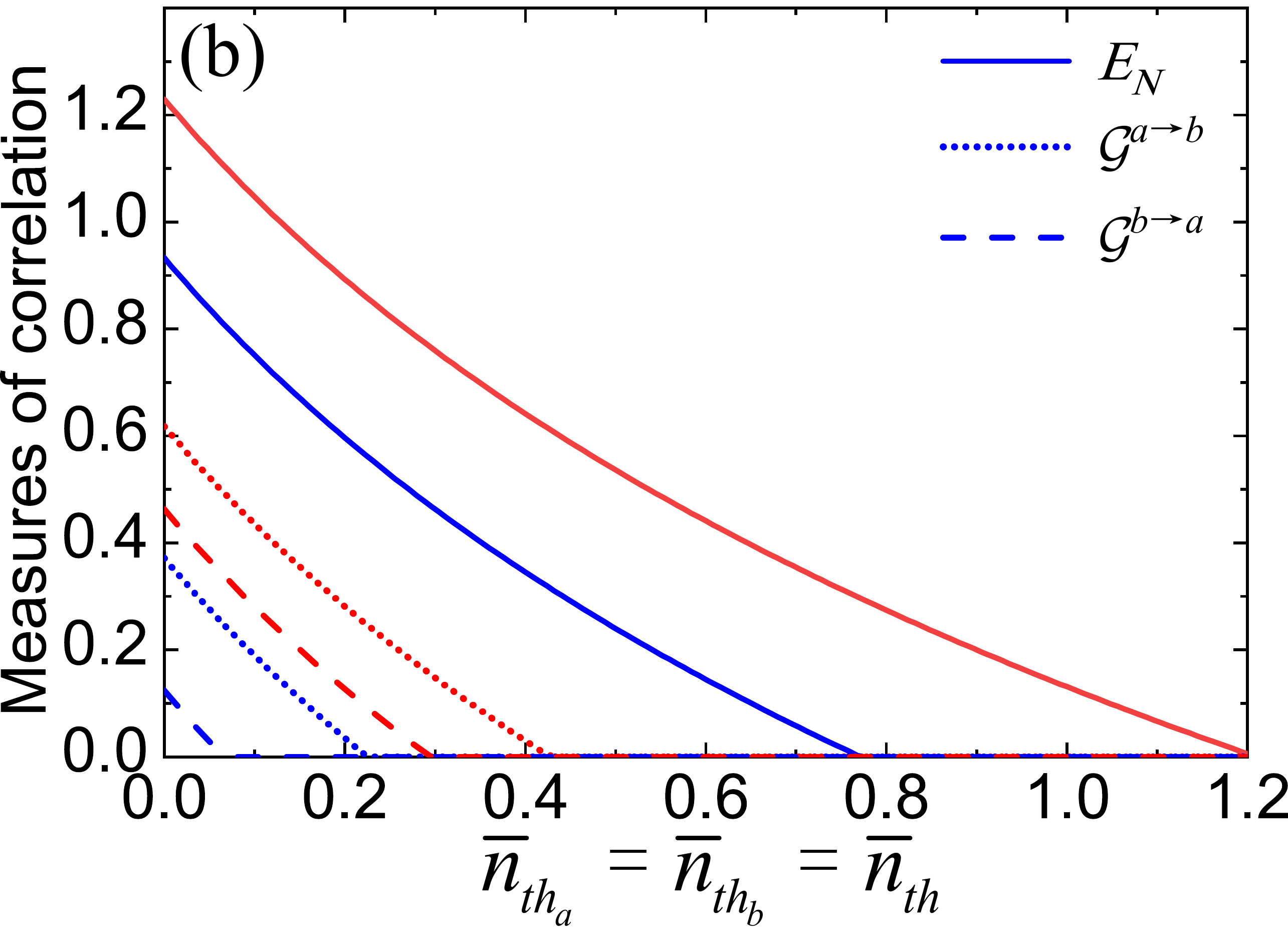}
\caption{The influence of thermal noise on the entanglement ($E_N$, solid) and steering ($\mathcal{G}^{a\rightarrow b}$ dotted and $\mathcal{G}^{b\rightarrow a}$ dashed) between modes $a$ and $b$ at phase $\phi=3\pi/2$ when (a) $\bar{n}_{th}$ only exists on the intermediate damping mode $c$, (b) $\bar{n}_{th_a}=\bar{n}_{th_b}=\bar{n}_{th}$. Blue curves indicate the results achieved only by the indirect interaction path where $\lambda=0$, and correspondingly red curves represent the enhanced correlations via interference channels when $\lambda=1.5\kappa$. Other parameters are $g_a=8.3\kappa,\ g_b=10\kappa,\gamma_c=5\kappa$.}
\label{bbr_nth}
\end{figure}

However, for two modes possessing completely symmetric decoherence properties we show in Fig.~\ref{bbr_coherent_lambda0} that by applying the interfering channels one can still create asymmetric and directional steering. Meanwhile, the degree of correlations is enhanced even at phase $\phi=\pi/2$ [Fig.~\ref{bbr_coherent_lambda0}(a)] that induces destructive interference, and the enhancement is more remarkable at phase $\phi=3\pi/2$ due to constructive interference  [Fig.~\ref{bbr_coherent_lambda0}(b)]. The blue curves in two plots are the entanglement achieved by the direct interaction path, while steering in both directions cannot be produced ($\mathcal{G}=0$ not plotted in the figure for clearity). When we include the other indirect coupling path, the entanglement (black dash-dotted) is remarkably enhanced, and asymmetric steering in the direction $\mathcal{G}^{a\rightarrow b}$ (red dotted) and in the opposite direction $\mathcal{G}^{b\rightarrow a}$ (red dashed) appear. Note that with destructive interference shown in Fig.~\ref{bbr_coherent_lambda0}(a), although correlations are reduced when $\lambda$ increases, it is useful to achieve one-way steering $\mathcal{G}^{b\rightarrow a}=0$ and $\mathcal{G}^{a\rightarrow b}>0$. With constructive interference shown in Fig.~\ref{bbr_coherent_lambda0}(b), the enhancement of entanglement and steering becomes notable,  and two-way asymmetric steering occurs for all steady-state solution. This observation consists with that shown in Fig.~\ref{bbr_dissipation_gamma} and the reason why $\mathcal{G}^{a\rightarrow b}>\mathcal{G}^{b\rightarrow a}$ has been explained there. Moreover, the parameter range to achieve steady-state solution is considerably broadened via interference channel. The stability condition given in Eq.~(\ref{stability condition}) at phase $\phi=\frac{\pi}{2}+n\pi\ (n \in \mathbb{Z})$ requires $\lambda^2<\text{min}\{\kappa_a\kappa_b+(g_b^2\kappa_a-g_a^2\kappa_b)/\gamma_c, \ \kappa_a\kappa_b+\gamma_c(\kappa_a+\kappa_b+\gamma_c)+[g_b^2(\kappa_b+\gamma_c)-g_a^2(\kappa_a+\gamma_c)]/(\kappa_a+\kappa_b)\}$, for the parameters studied here, the threshold is larger than $\kappa_a\kappa_b$ required by the direct interaction path only, that is to say, the system is more stable by applying our interference approach.

So far we have shown how the entanglement and steering can be enhanced and the direction of steering can be controlled via interference effects when three modes are not thermal excited. In practical implementation using cavity optomechanical systems, the third intermediate mode could be a mechanical mode with nonzero thermal phonons, so we show the influence of thermal noise $\bar{n}_{th}$ only existing in mode $c$ on the correlations at phase $\phi=3\pi/2$ in Fig.~\ref{bbr_nth}(a). In general, thermal noise is detrimental to correlations because it causes decoherence, however, entanglement (red solid) and steering (red dotted and dashed) created by the interference channels are more robust against thermal noise comparing with that achieved only by the indirect interaction path (correspondingly blue curves). Even when all three modes are in thermal excited states, it shows again in Fig.~\ref{bbr_nth}(b) that the entanglement and asymmetric steering created by the interference channels (red curves) are more robust against thermal noise than that achieved by only applying the indirect interaction path (blue curves).

\subsection{The enhancement of entanglement and steering between other two modes in the loop}
\begin{figure}[htbp!]
\includegraphics[width=0.8\columnwidth]
{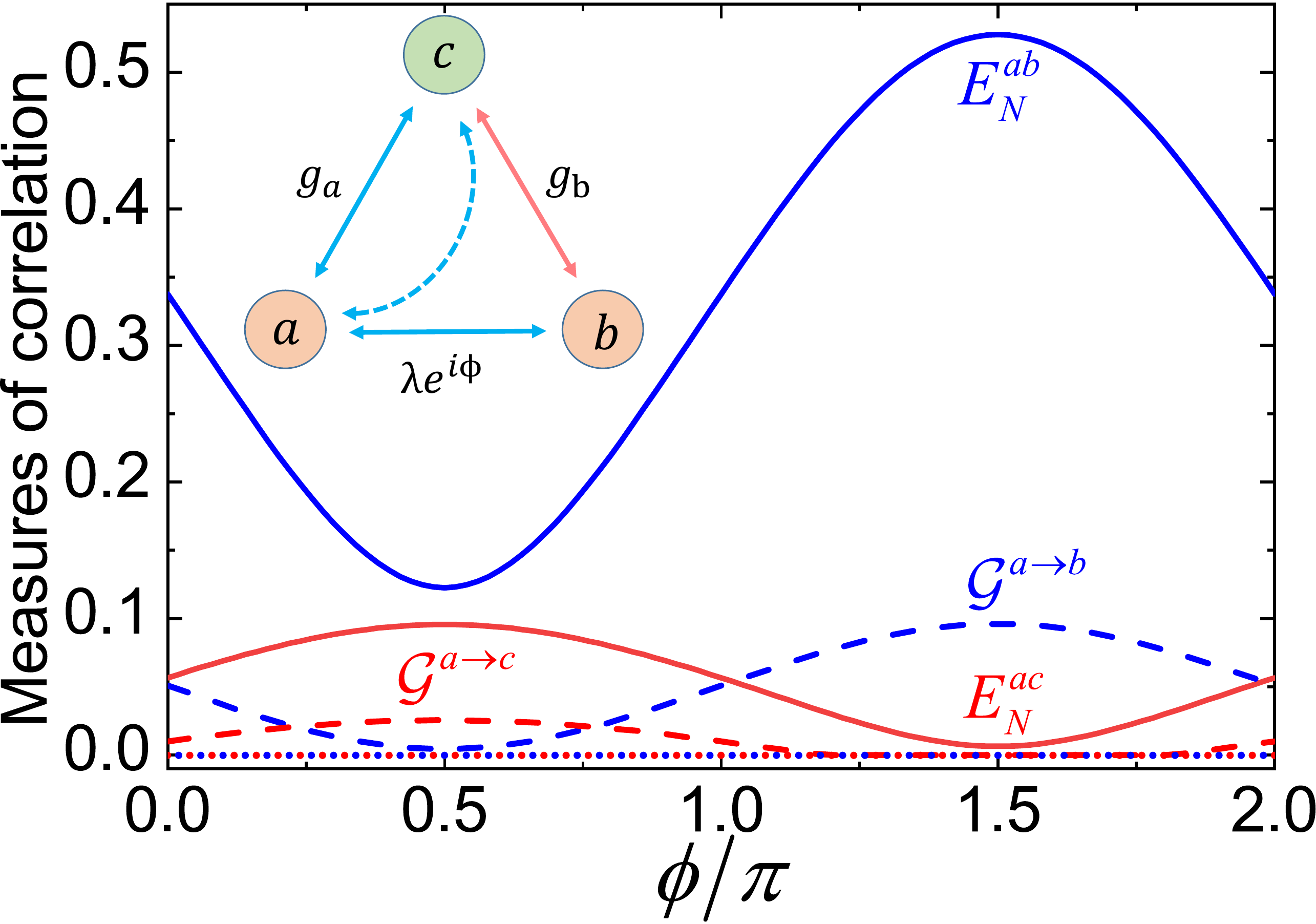}
\caption{The phase-dependent behavior of steady-state entanglement and steering between pair $a,b$ (blue curves) and between pair $a,c$ (red curves). When entanglement ($E^{ab}_{N}>0$, blue solid) and one-way steering ($\mathcal{G}^{a\rightarrow b}>0$, blue dashed, and $\mathcal{G}^{b\rightarrow a}=0$, blue dotted) between modes $a$ and $b$ reach a maximum, the entanglement ($E^{ac}_{N}>0$, red solid) and one-way steering ($\mathcal{G}^{a\rightarrow c}>0$, red dashed, and $\mathcal{G}^{c\rightarrow a}=0$, red dotted) between modes $a$ and $c$ are at a minimum. Other parameters are $\kappa_a=\kappa_b=\kappa,\ g_a=3.2\kappa,\ g_b=5\kappa,\ \lambda=0.4\kappa,\ \gamma_c=15\kappa,\ \bar{n}_{th}=0$.}
\label{phase_two pairs}
\end{figure}

We focused on the enhancement of entanglement and the control of the asymmetric and directional steering via interference effects between modes $a$ and $b$ in the previous discussion, here, we want to point out that this could be also realized between modes $a$ and $c$, which means the method is general. As shown by the inset plot in Fig.~\ref{phase_two pairs}, there are two interaction paths to create entanglement between modes $a$ and $c$, one is by the direct two-mode squeezing interaction with coupling strength $g_a$, and the other one is by swapping the entanglement shared by $a$ and $b$ to pair $a$ and $c$ via the beam-splitter-type interaction between modes $b$ and $c$. Two physical interaction paths superpose with relative phase and create interfering channels. The phase-dependent behavior of the entanglement ($E^{ac}_N$ red solid) and one-way steering between modes $a$ and $c$ ($\mathcal{G}^{a\rightarrow c}>0$ red dashed, $\mathcal{G}^{c\rightarrow a}=0$ red dotted) is illustrated in Fig.~\ref{phase_two pairs}. We also notice that when the entanglement and steering between modes $a$ and $c$ reaches a maximum (minimum) of the interference pattern at where the corresponding correlations between modes $a$ and $b$ are a minimum (maximum), indicating that there exists a competition to distribute entanglement and steering between different pairs. This can be understood by the monogamy constraints for distributing correlations among multi parties.

\section{Conclusion}\label{SUMMARY}
In summary, we propose an active method to generate and control steady-state entanglement and asymmetric steering between two identical modes by intriguing constructive or destructive quantum interference effects in a general three-mode system with closed-loop in coupling. The direct interaction path between two modes and the indirect interaction path induced by both coupling to the intermediate damping mode superpose and act as a interfering channel with a tunable phase, which can create phase-dependent correlations. We show that the interference effects can not only enhance the degree of the entanglement and steerability, but also produce asymmetric steering when two modes possess completely symmetric decoherence properties. Instead of introducing different amount of losses or noises to subsystems, this provides inspiration for manipulating the direction of the asymmetric quantum steering with enhanced steerability. In addition, the entangled steady states created by the interference channel are more robust against thermal noises. Furthermore, using the interfering channels to entangle the other pair of modes in the loop is also discussed, and the opposite phase-dependent behavior indicates the monogamy constraints for distributing entanglement and steering among multipartite. This work opens up new perspectives for the experimental production and applications of EPR steering as a precious resource for secure quantum communication technologies.

\begin{acknowledgements}
We acknowledge illuminating discussions with Y. Li and H.Y.Yuan. This work is supported by the National Key R\&D Program of China (Grants No. 2018YFB1107200 and No. 2016YFA0301302) and the National Natural Science Foundation of China (Grants No. 11622428, No. 61475007, and No. 61475006). Q.He thanks the Beijing Computational Science Research Center for their hospitality.
\end{acknowledgements}

\appendix\label{APPENDIX}
\section{The steady-state solution}\label{solution}

To signify the entanglement and Gaussian steering between modes $a$ and $b$ by the measure of logarithmic negativity $E_N$ and the quantities $\mathcal{G}^{a\rightarrow b}$ and $\mathcal{G}^{b\rightarrow a}$, we need to give out the steady-state solution of covariance matrix $V$ 
\begin{equation}
V=\left(
\begin{array}{cccc}
V(X_a) & 0 & V(X_a,X_b) & V(X_a,P_b) \\
0 & V(P_a) & V(P_a,X_b) & V(P_a,P_b) \\
V(X_b,X_a) & V(X_b,P_a) & V(X_b)  & 0 \\
V(P_b,X_a) &V(P_b,P_a) & 0 & V(P_b) 
\end{array}
\right),
\end{equation}
where the variances are defined $V(A)=\langle A^2\rangle-\langle A \rangle^2$ and $V(A,B)=\langle AB +BA\rangle/2-\langle A\rangle\langle B\rangle$. By solving the Lyapunov equation $MV+V M^{T}=-D$~\cite{Genes2008,Barzanjeh2011}, where the drift matrix $M$ and the diffusion matrix $D$ are given in main text, we can get the steady-state solution of all components in $V$.

For Gaussian systems, $\langle X_a\rangle=\langle P_a\rangle=\langle X_b\rangle=\langle P_b\rangle=0$, such that $\langle a^\dagger a\rangle=[V(X_a)+V(P_a)-1]/2$,  $\langle b^\dagger b\rangle=[V(X_b)+V(P_b)-1]/2$, and $\langle ab\rangle=[V(X_a,X_b)-V(P_a,P_b)+iV(X_a,P_b)+iV(P_a,X_b)]/2$. Specifically, for $\phi=\frac{\pi}{2}+n\pi\ (n \in \mathbb{Z})$, $V(X_a,P_b)=V(P_a,X_b)=0$, then the steady-state solutions of mode populations and correlation term are obtained as follows
\begin{widetext}
\begin{eqnarray}
\langle a^\dagger a\rangle=&&\Big\{g_a^2g_b^2(\kappa_a+\kappa_b+\gamma_c)\kappa_b+g_a^2\gamma_c(\bar{n}_{th}+1)[\kappa_ag_b^2-\kappa_bg_a^2+\kappa_b(\kappa_b+\gamma_c)(\kappa_a+\kappa_b)]\nonumber\\
&&\ \ \ +\lambda^2\{\kappa_b[\kappa_ag_b^2-\kappa_bg_a^2+\gamma_c(\kappa_a+\gamma_c)(\kappa_b+\gamma_c)]+\gamma_c(\bar{n}_{th}+1)[(\kappa_a+\kappa_b+\gamma_c)g_b^2-\gamma_cg_a^2]\}\nonumber\\
&&\ \ \ \ -\lambda^4\kappa_b\gamma_c-2\lambda\bar{n}_{th}\kappa_b\gamma_cg_ag_b(\kappa_a+\kappa_b+\gamma_c)\sin\phi\Big\}/De,
\nonumber
\end{eqnarray}
\begin{eqnarray}
\langle b^\dagger b\rangle=&&\Big\{g_a^2g_b^2(\kappa_a+\kappa_b+\gamma_c)\kappa_a+g_b^2\gamma_c\bar{n}_{th}[\kappa_ag_b^2-\kappa_bg_a^2+\kappa_a(\kappa_a+\gamma_c)(\kappa_a+\kappa_b)]\nonumber\\
&&\ \ \ +\lambda^2\{\kappa_a[\kappa_ag_b^2-\kappa_bg_a^2+\gamma_c(\kappa_a+\gamma_c)(\kappa_b+\gamma_c)]+\gamma_c\bar{n}_{th}[(\kappa_a+\kappa_b+\gamma_c)g_a^2-\gamma_cg_b^2]\}\nonumber\\
&&\ \ \ \ -\lambda^4\kappa_a\gamma_c-2\lambda(\bar{n}_{th}+1)\kappa_a\gamma_cg_ag_b(\kappa_a+\kappa_b+\gamma_c)\sin{\phi}\Big\}/De,
\nonumber
\end{eqnarray}
\begin{eqnarray}
\langle ab\rangle=&&-\Big\{g_ag_b\kappa_a[\kappa_bg_a^2+(\kappa_b+\gamma_c)(g_b^2+\kappa_b\gamma_c)]+g_ag_b\gamma_c\bar{n}_{th}[\kappa_ag_b^2-\kappa_bg_a^2+\kappa_a\kappa_b(\kappa_a+\kappa_b+2\gamma_c)]\nonumber\\
&&\ \ \ \ +\lambda\sin{\phi}[\kappa_b^2\kappa_ag_a^2-\kappa_a(g_b^2+\kappa_b\gamma_c)(\kappa_a+\gamma_c)(\kappa_b+\gamma_c)-\gamma_c\bar{n}_{th}(\kappa_a+\kappa_b+\gamma_c)(\kappa_ag_b^2+\kappa_bg_a^2)]\nonumber\\
&&\ \ \ \ \ \ +\lambda^3\sin{\phi}\kappa_a\kappa_b\gamma_c+\lambda^2g_ag_b\gamma_c[\kappa_a(\bar{n}_{th}+1)+\kappa_b\bar{n}_{th}]\Big\}/De,  \nonumber
\end{eqnarray}
\begin{eqnarray}\label{correlation_term}
De=&(\kappa_bg_a^2-\kappa_ag_b^2-\gamma_c\kappa_a\kappa_b+\gamma_c\lambda^2)\big\{(\kappa_a+\gamma_c)g_a^2-(\kappa_b+\gamma_c)g_b^2+[\lambda^2-(\kappa_a+\gamma_c)(\kappa_b+\gamma_c)](\kappa_a+\kappa_b)\big\},
\end{eqnarray}
\end{widetext}
where $\sin\phi=+1$ and $\sin\phi=-1$ for destructive and constructive interference when $\phi=(2n+1/2)\pi$ and $\phi=(2n+3/2)\pi$, respectively.

\section{The derivation of the entanglement and steering conditions given in Eq.~(\ref{G21})}\label{derivation}

In this part we will simply derive the equivalence between the measures of entanglement $E_N>0$ and Gaussian steering $\mathcal{G}^{a\rightarrow b}>0$ and $\mathcal{G}^{b\rightarrow a}>0$ and the inequalities given in Eq.~(\ref{G21}).

In the present scheme, we have $V(X_a)=V(P_a)=n_a$, $V(X_b)=V(P_b)=n_b$, $V(X_a,X_b)=-V(P_a,P_b)=c_1$, $V(X_a,P_b)=V(P_a,X_b)=c_2$, and therefore the CM of modes $a$ and $b$ at any phase $\phi$ can be expressed as
\begin{equation}
V=\left(
\begin{array}{cccc}
n_a & 0 & c_1 & c_2 \\
0 & n_a & c_2 & -c_1 \\
c_1 & c_2 & n_b & 0 \\
c_2 & -c_1 & 0 & n_b
\end{array}
\right),
\end{equation}
The logarithmic negativity $E_N>0$ to confirm entanglement requires $\eta^-\equiv\sqrt{\Sigma(V)-[\Sigma(V)^2-4\det{V}]^{1/2}}/\sqrt{2}<1/2$, which is equivalent to
\begin{equation}
[c_1^2+c_2^2-(n_a+\frac{1}{2})(n_b+\frac{1}{2})][c_1^2+c_2^2-(n_a-\frac{1}{2})(n_b-\frac{1}{2})]<0.
\end{equation}
Noting that the Cauchy-Schwarz inequality implies that $|\langle ab \rangle|^2\leq \langle a^\dagger a\rangle(\langle b^\dagger b\rangle+1)$, i.e., $c_1^2+c_2^2\leq(n_a-1/2)(n_b+1/2)$, we find that $c_1^2+c_2^2<(n_a+1/2)(n_b+1/2)$ must be true, and then the above entanglement condition reduces to
\begin{equation}
|\langle ab\rangle|>\sqrt{\langle a^\dagger a\rangle \langle b^\dagger b\rangle}.
\end{equation}
This criterion was directly provided by Hillery and Zubairy~\cite{Hillery2006} for two-mode states by examining uncertainty relations, and the derivation from the logarithmic negativity in the Gaussian regime (with the condition $\langle a^2\rangle=\langle b^2\rangle=\langle a^\dagger b\rangle=0$) has been shown in Ref.~\cite{Tan2015}.

In analogy to what was done for entanglement, we can also derive the measure of Gaussian steering $\mathcal{G}$ in terms of population and correlations. The steering from mode $a$ to mode $b$ occurs iff $\mathcal{G}^{a\rightarrow b}>0$, which is equivalent to
\begin{equation}
[c_1^2+c_2^2-n_a(n_b+\frac{1}{2})][c_1^2+c_2^2-n_a(n_b-\frac{1}{2})]<0.
\end{equation}
The Cauchy-Schwarz inequality implies that $c_1^2+c_2^2\leq(n_a-1/2)(n_b+1/2)<n_a(n_b+1/2)$, then the above steering condition reduces to
\begin{equation}
|\langle ab\rangle|>\sqrt{\langle b^\dagger b\rangle(\langle a^\dagger a\rangle+1/2)}.
\end{equation}
Similarly, the condition $\mathcal{G}^{b\rightarrow a}>0$ reduces to
\begin{equation}
|\langle ab\rangle|>\sqrt{\langle a^\dagger a\rangle(\langle b^\dagger b\rangle+1/2)}.
\end{equation}
Note that this criteria for bipartite and multipartite steering has been directly developed in Ref.~\cite{Cavalcanti2011}, and can be also derived from Reid criterion in terms of variances in the Gaussian regime (with the condition $\langle a^2\rangle=\langle b^2\rangle=\langle a^\dagger b\rangle=0$)~\cite{Tan2015}. 

\bibliography{Control_Directional_Correlation_via_Interference}

\begin{thebibliography}{66}%
\makeatletter
\providecommand \@ifxundefined [1]{%
 \@ifx{#1\undefined}
}%
\providecommand \@ifnum [1]{%
 \ifnum #1\expandafter \@firstoftwo
 \else \expandafter \@secondoftwo
 \fi
}%
\providecommand \@ifx [1]{%
 \ifx #1\expandafter \@firstoftwo
 \else \expandafter \@secondoftwo
 \fi
}%
\providecommand \natexlab [1]{#1}%
\providecommand \enquote  [1]{``#1''}%
\providecommand \bibnamefont  [1]{#1}%
\providecommand \bibfnamefont [1]{#1}%
\providecommand \citenamefont [1]{#1}%
\providecommand \href@noop [0]{\@secondoftwo}%
\providecommand \href [0]{\begingroup \@sanitize@url \@href}%
\providecommand \@href[1]{\@@startlink{#1}\@@href}%
\providecommand \@@href[1]{\endgroup#1\@@endlink}%
\providecommand \@sanitize@url [0]{\catcode `\\12\catcode `\$12\catcode
  `\&12\catcode `\#12\catcode `\^12\catcode `\_12\catcode `\%12\relax}%
\providecommand \@@startlink[1]{}%
\providecommand \@@endlink[0]{}%
\providecommand \url  [0]{\begingroup\@sanitize@url \@url }%
\providecommand \@url [1]{\endgroup\@href {#1}{\urlprefix }}%
\providecommand \urlprefix  [0]{URL }%
\providecommand \Eprint [0]{\href }%
\providecommand \doibase [0]{http://dx.doi.org/}%
\providecommand \selectlanguage [0]{\@gobble}%
\providecommand \bibinfo  [0]{\@secondoftwo}%
\providecommand \bibfield  [0]{\@secondoftwo}%
\providecommand \translation [1]{[#1]}%
\providecommand \BibitemOpen [0]{}%
\providecommand \bibitemStop [0]{}%
\providecommand \bibitemNoStop [0]{.\EOS\space}%
\providecommand \EOS [0]{\spacefactor3000\relax}%
\providecommand \BibitemShut  [1]{\csname bibitem#1\endcsname}%
\let\auto@bib@innerbib\@empty
\bibitem [{\citenamefont {Horodecki}\ \emph {et~al.}(2009)\citenamefont
  {Horodecki}, \citenamefont {Horodecki}, \citenamefont {Horodecki},\ and\
  \citenamefont {Horodecki}}]{Horodecki2009}%
  \BibitemOpen
  \bibfield  {author} {\bibinfo {author} {\bibfnamefont {R.}~\bibnamefont
  {Horodecki}}, \bibinfo {author} {\bibfnamefont {P.}~\bibnamefont
  {Horodecki}}, \bibinfo {author} {\bibfnamefont {M.}~\bibnamefont
  {Horodecki}}, \ and\ \bibinfo {author} {\bibfnamefont {K.}~\bibnamefont
  {Horodecki}},\ }\href {\doibase 10.1103/RevModPhys.81.865} {\bibfield
  {journal} {\bibinfo  {journal} {Rev. Mod. Phys.}\ }\textbf {\bibinfo {volume}
  {81}},\ \bibinfo {pages} {865} (\bibinfo {year} {2009})}\BibitemShut
  {NoStop}%
\bibitem [{\citenamefont {Reid}\ \emph {et~al.}(2009)\citenamefont {Reid},
  \citenamefont {Drummond}, \citenamefont {Bowen}, \citenamefont {Cavalcanti},
  \citenamefont {Lam}, \citenamefont {Bachor}, \citenamefont {Andersen},\ and\
  \citenamefont {Leuchs}}]{Reid2009}%
  \BibitemOpen
  \bibfield  {author} {\bibinfo {author} {\bibfnamefont {M.~D.}\ \bibnamefont
  {Reid}}, \bibinfo {author} {\bibfnamefont {P.~D.}\ \bibnamefont {Drummond}},
  \bibinfo {author} {\bibfnamefont {W.~P.}\ \bibnamefont {Bowen}}, \bibinfo
  {author} {\bibfnamefont {E.~G.}\ \bibnamefont {Cavalcanti}}, \bibinfo
  {author} {\bibfnamefont {P.~K.}\ \bibnamefont {Lam}}, \bibinfo {author}
  {\bibfnamefont {H.~A.}\ \bibnamefont {Bachor}}, \bibinfo {author}
  {\bibfnamefont {U.~L.}\ \bibnamefont {Andersen}}, \ and\ \bibinfo {author}
  {\bibfnamefont {G.}~\bibnamefont {Leuchs}},\ }\href {\doibase
  10.1103/RevModPhys.81.1727} {\bibfield  {journal} {\bibinfo  {journal} {Rev.
  Mod. Phys.}\ }\textbf {\bibinfo {volume} {81}},\ \bibinfo {pages} {1727}
  (\bibinfo {year} {2009})}\BibitemShut {NoStop}%
\bibitem [{\citenamefont {Cavalcanti}\ and\ \citenamefont
  {Skrzypczyk}(2017)}]{PhysRep2017}%
  \BibitemOpen
  \bibfield  {author} {\bibinfo {author} {\bibfnamefont {D.}~\bibnamefont
  {Cavalcanti}}\ and\ \bibinfo {author} {\bibfnamefont {P.}~\bibnamefont
  {Skrzypczyk}},\ }\href {\doibase 10.1088/1361-6633/80/2/024001} {\bibfield
  {journal} {\bibinfo  {journal} {Rep. Prog. Phys.}\ }\textbf {\bibinfo
  {volume} {80}},\ \bibinfo {pages} {024001} (\bibinfo {year}
  {2017})}\BibitemShut {NoStop}%
\bibitem [{\citenamefont {Schr{\"{o}}dinger}(1935)}]{Schrodinger1935}%
  \BibitemOpen
  \bibfield  {author} {\bibinfo {author} {\bibfnamefont {E.}~\bibnamefont
  {Schr{\"{o}}dinger}},\ }\href@noop {} {\bibfield  {journal} {\bibinfo
  {journal} {Proc. Cambridge Philos. Soc.}\ }\textbf {\bibinfo {volume} {31}},\
  \bibinfo {pages} {555} (\bibinfo {year} {1935})}\BibitemShut {NoStop}%
\bibitem [{\citenamefont {Einstein}\ \emph {et~al.}(1935)\citenamefont
  {Einstein}, \citenamefont {Podolsky},\ and\ \citenamefont
  {Rosen}}]{Einstein1935}%
  \BibitemOpen
  \bibfield  {author} {\bibinfo {author} {\bibfnamefont {A.}~\bibnamefont
  {Einstein}}, \bibinfo {author} {\bibfnamefont {B.}~\bibnamefont {Podolsky}},
  \ and\ \bibinfo {author} {\bibfnamefont {N.}~\bibnamefont {Rosen}},\ }\href
  {\doibase 10.1007/s10701-010-9411-9} {\bibfield  {journal} {\bibinfo
  {journal} {Phys. Rev.}\ }\textbf {\bibinfo {volume} {47}},\ \bibinfo {pages}
  {777} (\bibinfo {year} {1935})}\BibitemShut {NoStop}%
\bibitem [{\citenamefont {Reid}(1989)}]{Reid1989}%
  \BibitemOpen
  \bibfield  {author} {\bibinfo {author} {\bibfnamefont {M.~D.}\ \bibnamefont
  {Reid}},\ }\href {\doibase 10.1103/PhysRevA.40.913} {\bibfield  {journal}
  {\bibinfo  {journal} {Phys. Rev. A}\ }\textbf {\bibinfo {volume} {40}},\
  \bibinfo {pages} {913} (\bibinfo {year} {1989})}\BibitemShut {NoStop}%
\bibitem [{\citenamefont {Wiseman}\ \emph {et~al.}(2007)\citenamefont
  {Wiseman}, \citenamefont {Jones},\ and\ \citenamefont
  {Doherty}}]{Wiseman2007}%
  \BibitemOpen
  \bibfield  {author} {\bibinfo {author} {\bibfnamefont {H.~M.}\ \bibnamefont
  {Wiseman}}, \bibinfo {author} {\bibfnamefont {S.~J.}\ \bibnamefont {Jones}},
  \ and\ \bibinfo {author} {\bibfnamefont {A.~C.}\ \bibnamefont {Doherty}},\
  }\href {\doibase 10.1103/PhysRevLett.98.140402} {\bibfield  {journal}
  {\bibinfo  {journal} {Phys. Rev. Lett.}\ }\textbf {\bibinfo {volume} {98}},\
  \bibinfo {pages} {140402} (\bibinfo {year} {2007})}\BibitemShut {NoStop}%
\bibitem [{\citenamefont {Jones}\ \emph {et~al.}(2007)\citenamefont {Jones},
  \citenamefont {Wiseman},\ and\ \citenamefont {Doherty}}]{Jones2007}%
  \BibitemOpen
  \bibfield  {author} {\bibinfo {author} {\bibfnamefont {S.~J.}\ \bibnamefont
  {Jones}}, \bibinfo {author} {\bibfnamefont {H.~M.}\ \bibnamefont {Wiseman}},
  \ and\ \bibinfo {author} {\bibfnamefont {A.~C.}\ \bibnamefont {Doherty}},\
  }\href {\doibase 10.1103/PhysRevA.76.052116} {\bibfield  {journal} {\bibinfo
  {journal} {Phys. Rev. A}\ }\textbf {\bibinfo {volume} {76}},\ \bibinfo
  {pages} {052116} (\bibinfo {year} {2007})}\BibitemShut {NoStop}%
\bibitem [{\citenamefont {Cavalcanti}\ \emph {et~al.}(2009)\citenamefont
  {Cavalcanti}, \citenamefont {Jones}, \citenamefont {Wiseman},\ and\
  \citenamefont {Reid}}]{Cavalcanti2009}%
  \BibitemOpen
  \bibfield  {author} {\bibinfo {author} {\bibfnamefont {E.~G.}\ \bibnamefont
  {Cavalcanti}}, \bibinfo {author} {\bibfnamefont {S.~J.}\ \bibnamefont
  {Jones}}, \bibinfo {author} {\bibfnamefont {H.~M.}\ \bibnamefont {Wiseman}},
  \ and\ \bibinfo {author} {\bibfnamefont {M.~D.}\ \bibnamefont {Reid}},\
  }\href {\doibase 10.1103/PhysRevA.80.032112} {\bibfield  {journal} {\bibinfo
  {journal} {Phys. Rev. A}\ }\textbf {\bibinfo {volume} {80}},\ \bibinfo
  {pages} {032112} (\bibinfo {year} {2009})}\BibitemShut {NoStop}%
\bibitem [{\citenamefont {Opanchuk}\ \emph {et~al.}(2014)\citenamefont
  {Opanchuk}, \citenamefont {Arnaud},\ and\ \citenamefont
  {Reid}}]{Opanchuk2014}%
  \BibitemOpen
  \bibfield  {author} {\bibinfo {author} {\bibfnamefont {B.}~\bibnamefont
  {Opanchuk}}, \bibinfo {author} {\bibfnamefont {L.}~\bibnamefont {Arnaud}}, \
  and\ \bibinfo {author} {\bibfnamefont {M.~D.}\ \bibnamefont {Reid}},\ }\href
  {\doibase 10.1103/PhysRevA.89.062101} {\bibfield  {journal} {\bibinfo
  {journal} {Phys. Rev. A}\ }\textbf {\bibinfo {volume} {89}},\ \bibinfo
  {pages} {062101} (\bibinfo {year} {2014})}\BibitemShut {NoStop}%
\bibitem [{\citenamefont {Branciard}\ \emph {et~al.}(2012)\citenamefont
  {Branciard}, \citenamefont {Cavalcanti}, \citenamefont {Walborn},
  \citenamefont {Scarani},\ and\ \citenamefont {Wiseman}}]{Branciard2012}%
  \BibitemOpen
  \bibfield  {author} {\bibinfo {author} {\bibfnamefont {C.}~\bibnamefont
  {Branciard}}, \bibinfo {author} {\bibfnamefont {E.~G.}\ \bibnamefont
  {Cavalcanti}}, \bibinfo {author} {\bibfnamefont {S.~P.}\ \bibnamefont
  {Walborn}}, \bibinfo {author} {\bibfnamefont {V.}~\bibnamefont {Scarani}}, \
  and\ \bibinfo {author} {\bibfnamefont {H.~M.}\ \bibnamefont {Wiseman}},\
  }\href {\doibase 10.1103/PhysRevA.85.010301} {\bibfield  {journal} {\bibinfo
  {journal} {Phys. Rev. A}\ }\textbf {\bibinfo {volume} {85}},\ \bibinfo
  {pages} {010301(R)} (\bibinfo {year} {2012})}\BibitemShut {NoStop}%
\bibitem [{\citenamefont {Gehring}\ \emph {et~al.}(2015)\citenamefont
  {Gehring}, \citenamefont {H{\"{a}}ndchen}, \citenamefont {Duhme},
  \citenamefont {Furrer}, \citenamefont {Franz}, \citenamefont {Pacher},
  \citenamefont {Werner},\ and\ \citenamefont {Schnabel}}]{Gehring2015}%
  \BibitemOpen
  \bibfield  {author} {\bibinfo {author} {\bibfnamefont {T.}~\bibnamefont
  {Gehring}}, \bibinfo {author} {\bibfnamefont {V.}~\bibnamefont
  {H{\"{a}}ndchen}}, \bibinfo {author} {\bibfnamefont {J.}~\bibnamefont
  {Duhme}}, \bibinfo {author} {\bibfnamefont {F.}~\bibnamefont {Furrer}},
  \bibinfo {author} {\bibfnamefont {T.}~\bibnamefont {Franz}}, \bibinfo
  {author} {\bibfnamefont {C.}~\bibnamefont {Pacher}}, \bibinfo {author}
  {\bibfnamefont {R.~F.}\ \bibnamefont {Werner}}, \ and\ \bibinfo {author}
  {\bibfnamefont {R.}~\bibnamefont {Schnabel}},\ }\href {\doibase
  10.1038/ncomms9795} {\bibfield  {journal} {\bibinfo  {journal} {Nat.
  Commun.}\ }\textbf {\bibinfo {volume} {6}},\ \bibinfo {pages} {8795}
  (\bibinfo {year} {2015})}\BibitemShut {NoStop}%
\bibitem [{\citenamefont {Walk}\ \emph {et~al.}(2016)\citenamefont {Walk},
  \citenamefont {Hosseni}, \citenamefont {Geng}, \citenamefont {Thearle},
  \citenamefont {Haw}, \citenamefont {Armstrong}, \citenamefont {Assad},
  \citenamefont {Janousek}, \citenamefont {Ralph}, \citenamefont {Symul},
  \citenamefont {Wiseman},\ and\ \citenamefont {Lam}}]{Walk2016}%
  \BibitemOpen
  \bibfield  {author} {\bibinfo {author} {\bibfnamefont {N.}~\bibnamefont
  {Walk}}, \bibinfo {author} {\bibfnamefont {S.}~\bibnamefont {Hosseni}},
  \bibinfo {author} {\bibfnamefont {J.}~\bibnamefont {Geng}}, \bibinfo {author}
  {\bibfnamefont {O.}~\bibnamefont {Thearle}}, \bibinfo {author} {\bibfnamefont
  {J.~Y.}\ \bibnamefont {Haw}}, \bibinfo {author} {\bibfnamefont
  {S.}~\bibnamefont {Armstrong}}, \bibinfo {author} {\bibfnamefont {S.~M.}\
  \bibnamefont {Assad}}, \bibinfo {author} {\bibfnamefont {J.}~\bibnamefont
  {Janousek}}, \bibinfo {author} {\bibfnamefont {T.~C.}\ \bibnamefont {Ralph}},
  \bibinfo {author} {\bibfnamefont {T.}~\bibnamefont {Symul}}, \bibinfo
  {author} {\bibfnamefont {H.~M.}\ \bibnamefont {Wiseman}}, \ and\ \bibinfo
  {author} {\bibfnamefont {P.~K.}\ \bibnamefont {Lam}},\ }\href {\doibase
  10.1364/OPTICA.3.000634} {\bibfield  {journal} {\bibinfo  {journal} {Optica}\
  }\textbf {\bibinfo {volume} {3}},\ \bibinfo {pages} {634} (\bibinfo {year}
  {2016})}\BibitemShut {NoStop}%
\bibitem [{\citenamefont {Armstrong}\ \emph {et~al.}(2015)\citenamefont
  {Armstrong}, \citenamefont {Wang}, \citenamefont {Teh}, \citenamefont {Gong},
  \citenamefont {He}, \citenamefont {Janousek}, \citenamefont {Bachor},
  \citenamefont {Reid},\ and\ \citenamefont {Lam}}]{Armstrong2015}%
  \BibitemOpen
  \bibfield  {author} {\bibinfo {author} {\bibfnamefont {S.}~\bibnamefont
  {Armstrong}}, \bibinfo {author} {\bibfnamefont {M.}~\bibnamefont {Wang}},
  \bibinfo {author} {\bibfnamefont {R.~Y.}\ \bibnamefont {Teh}}, \bibinfo
  {author} {\bibfnamefont {Q.~H.}\ \bibnamefont {Gong}}, \bibinfo {author}
  {\bibfnamefont {Q.~Y.}\ \bibnamefont {He}}, \bibinfo {author} {\bibfnamefont
  {J.}~\bibnamefont {Janousek}}, \bibinfo {author} {\bibfnamefont {H.~A.}\
  \bibnamefont {Bachor}}, \bibinfo {author} {\bibfnamefont {M.~D.}\
  \bibnamefont {Reid}}, \ and\ \bibinfo {author} {\bibfnamefont {P.~K.}\
  \bibnamefont {Lam}},\ }\href {\doibase 10.1038/nphys3202} {\bibfield
  {journal} {\bibinfo  {journal} {Nat. Phys.}\ }\textbf {\bibinfo {volume}
  {11}},\ \bibinfo {pages} {167} (\bibinfo {year} {2015})}\BibitemShut
  {NoStop}%
\bibitem [{\citenamefont {Xiang}\ \emph {et~al.}(2017)\citenamefont {Xiang},
  \citenamefont {Kogias}, \citenamefont {Adesso},\ and\ \citenamefont
  {He}}]{Xiang2017}%
  \BibitemOpen
  \bibfield  {author} {\bibinfo {author} {\bibfnamefont {Y.}~\bibnamefont
  {Xiang}}, \bibinfo {author} {\bibfnamefont {I.}~\bibnamefont {Kogias}},
  \bibinfo {author} {\bibfnamefont {G.}~\bibnamefont {Adesso}}, \ and\ \bibinfo
  {author} {\bibfnamefont {Q.~Y.}\ \bibnamefont {He}},\ }\href {\doibase
  10.1103/PhysRevA.95.010101} {\bibfield  {journal} {\bibinfo  {journal} {Phys.
  Rev. A}\ }\textbf {\bibinfo {volume} {95}},\ \bibinfo {pages} {010101(R)}
  (\bibinfo {year} {2017})}\BibitemShut {NoStop}%
\bibitem [{\citenamefont {Kogias}\ \emph {et~al.}(2017)\citenamefont {Kogias},
  \citenamefont {Xiang}, \citenamefont {He},\ and\ \citenamefont
  {Adesso}}]{Kogias2017}%
  \BibitemOpen
  \bibfield  {author} {\bibinfo {author} {\bibfnamefont {I.}~\bibnamefont
  {Kogias}}, \bibinfo {author} {\bibfnamefont {Y.}~\bibnamefont {Xiang}},
  \bibinfo {author} {\bibfnamefont {Q.~Y.}\ \bibnamefont {He}}, \ and\ \bibinfo
  {author} {\bibfnamefont {G.}~\bibnamefont {Adesso}},\ }\href {\doibase
  10.1103/PhysRevA.95.012315} {\bibfield  {journal} {\bibinfo  {journal} {Phys.
  Rev. A}\ }\textbf {\bibinfo {volume} {95}},\ \bibinfo {pages} {012315}
  (\bibinfo {year} {2017})}\BibitemShut {NoStop}%
\bibitem [{\citenamefont {Li}\ \emph {et~al.}(2015)\citenamefont {Li},
  \citenamefont {Chen}, \citenamefont {Chen}, \citenamefont {Zhang},
  \citenamefont {Chen},\ and\ \citenamefont {Pan}}]{Li2015}%
  \BibitemOpen
  \bibfield  {author} {\bibinfo {author} {\bibfnamefont {C.~M.}\ \bibnamefont
  {Li}}, \bibinfo {author} {\bibfnamefont {K.}~\bibnamefont {Chen}}, \bibinfo
  {author} {\bibfnamefont {Y.~N.}\ \bibnamefont {Chen}}, \bibinfo {author}
  {\bibfnamefont {Q.}~\bibnamefont {Zhang}}, \bibinfo {author} {\bibfnamefont
  {Y.~A.}\ \bibnamefont {Chen}}, \ and\ \bibinfo {author} {\bibfnamefont
  {J.~W.}\ \bibnamefont {Pan}},\ }\href {\doibase
  10.1103/PhysRevLett.115.010402} {\bibfield  {journal} {\bibinfo  {journal}
  {Phys. Rev. Lett.}\ }\textbf {\bibinfo {volume} {115}},\ \bibinfo {pages}
  {010402} (\bibinfo {year} {2015})}\BibitemShut {NoStop}%
\bibitem [{\citenamefont {He}\ \emph {et~al.}(2015{\natexlab{a}})\citenamefont
  {He}, \citenamefont {Rosales-Z{\'{a}}rate}, \citenamefont {Adesso},\ and\
  \citenamefont {Reid}}]{He2015}%
  \BibitemOpen
  \bibfield  {author} {\bibinfo {author} {\bibfnamefont {Q.~Y.}\ \bibnamefont
  {He}}, \bibinfo {author} {\bibfnamefont {L.}~\bibnamefont
  {Rosales-Z{\'{a}}rate}}, \bibinfo {author} {\bibfnamefont {G.}~\bibnamefont
  {Adesso}}, \ and\ \bibinfo {author} {\bibfnamefont {M.~D.}\ \bibnamefont
  {Reid}},\ }\href {\doibase 10.1103/PhysRevLett.115.180502} {\bibfield
  {journal} {\bibinfo  {journal} {Phys. Rev. Lett.}\ }\textbf {\bibinfo
  {volume} {115}},\ \bibinfo {pages} {180502} (\bibinfo {year}
  {2015}{\natexlab{a}})}\BibitemShut {NoStop}%
\bibitem [{\citenamefont {Reid}(2013)}]{Reid2013}%
  \BibitemOpen
  \bibfield  {author} {\bibinfo {author} {\bibfnamefont {M.~D.}\ \bibnamefont
  {Reid}},\ }\href {\doibase 10.1103/PhysRevA.88.062338} {\bibfield  {journal}
  {\bibinfo  {journal} {Phys. Rev. A}\ }\textbf {\bibinfo {volume} {88}},\
  \bibinfo {pages} {062338} (\bibinfo {year} {2013})}\BibitemShut {NoStop}%
\bibitem [{\citenamefont {Chiu}\ \emph {et~al.}(2016)\citenamefont {Chiu},
  \citenamefont {Lambert}, \citenamefont {Liao}, \citenamefont {Nori},\ and\
  \citenamefont {Li}}]{CMLi2016}%
  \BibitemOpen
  \bibfield  {author} {\bibinfo {author} {\bibfnamefont {C.-Y.}\ \bibnamefont
  {Chiu}}, \bibinfo {author} {\bibfnamefont {N.}~\bibnamefont {Lambert}},
  \bibinfo {author} {\bibfnamefont {T.-L.}\ \bibnamefont {Liao}}, \bibinfo
  {author} {\bibfnamefont {F.}~\bibnamefont {Nori}}, \ and\ \bibinfo {author}
  {\bibfnamefont {C.-M.}\ \bibnamefont {Li}},\ }\href {\doibase
  10.1038/npjqi.2016.20} {\bibfield  {journal} {\bibinfo  {journal} {NPJ
  Quantum Information}\ }\textbf {\bibinfo {volume} {2}},\ \bibinfo {pages}
  {16020} (\bibinfo {year} {2016})}\BibitemShut {NoStop}%
\bibitem [{\citenamefont {Piani}\ and\ \citenamefont
  {Watrous}(2015)}]{Piani2015}%
  \BibitemOpen
  \bibfield  {author} {\bibinfo {author} {\bibfnamefont {M.}~\bibnamefont
  {Piani}}\ and\ \bibinfo {author} {\bibfnamefont {J.}~\bibnamefont
  {Watrous}},\ }\href {\doibase 10.1103/PhysRevLett.114.060404} {\bibfield
  {journal} {\bibinfo  {journal} {Phys. Rev. Lett.}\ }\textbf {\bibinfo
  {volume} {114}},\ \bibinfo {pages} {060604} (\bibinfo {year}
  {2015})}\BibitemShut {NoStop}%
\bibitem [{\citenamefont {Wagner}\ \emph {et~al.}(2008)\citenamefont {Wagner},
  \citenamefont {Janousek}, \citenamefont {Delaubert}, \citenamefont {Zou},
  \citenamefont {Harb}, \citenamefont {Treps}, \citenamefont {Morizur},
  \citenamefont {Lam},\ and\ \citenamefont {Bachor}}]{Wagner2008}%
  \BibitemOpen
  \bibfield  {author} {\bibinfo {author} {\bibfnamefont {K.}~\bibnamefont
  {Wagner}}, \bibinfo {author} {\bibfnamefont {J.}~\bibnamefont {Janousek}},
  \bibinfo {author} {\bibfnamefont {V.}~\bibnamefont {Delaubert}}, \bibinfo
  {author} {\bibfnamefont {H.}~\bibnamefont {Zou}}, \bibinfo {author}
  {\bibfnamefont {C.}~\bibnamefont {Harb}}, \bibinfo {author} {\bibfnamefont
  {N.}~\bibnamefont {Treps}}, \bibinfo {author} {\bibfnamefont {J.~F.}\
  \bibnamefont {Morizur}}, \bibinfo {author} {\bibfnamefont {P.~K.}\
  \bibnamefont {Lam}}, \ and\ \bibinfo {author} {\bibfnamefont {H.~A.}\
  \bibnamefont {Bachor}},\ }\href {\doibase 10.1126/science.1159663} {\bibfield
   {journal} {\bibinfo  {journal} {Science}\ }\textbf {\bibinfo {volume}
  {321}},\ \bibinfo {pages} {541} (\bibinfo {year} {2008})}\BibitemShut
  {NoStop}%
\bibitem [{\citenamefont {H{\"{a}}ndchen}\ \emph {et~al.}(2012)\citenamefont
  {H{\"{a}}ndchen}, \citenamefont {Eberle}, \citenamefont {Steinlechner},
  \citenamefont {Samblowski}, \citenamefont {Franz}, \citenamefont {Werner},\
  and\ \citenamefont {Schnabel}}]{Handchen2012}%
  \BibitemOpen
  \bibfield  {author} {\bibinfo {author} {\bibfnamefont {V.}~\bibnamefont
  {H{\"{a}}ndchen}}, \bibinfo {author} {\bibfnamefont {T.}~\bibnamefont
  {Eberle}}, \bibinfo {author} {\bibfnamefont {S.}~\bibnamefont
  {Steinlechner}}, \bibinfo {author} {\bibfnamefont {A.}~\bibnamefont
  {Samblowski}}, \bibinfo {author} {\bibfnamefont {T.}~\bibnamefont {Franz}},
  \bibinfo {author} {\bibfnamefont {R.~F.}\ \bibnamefont {Werner}}, \ and\
  \bibinfo {author} {\bibfnamefont {R.}~\bibnamefont {Schnabel}},\ }\href
  {\doibase 10.1038/nphoton.2012.202} {\bibfield  {journal} {\bibinfo
  {journal} {Nat. Photonics}\ }\textbf {\bibinfo {volume} {6}},\ \bibinfo
  {pages} {596} (\bibinfo {year} {2012})}\BibitemShut {NoStop}%
\bibitem [{\citenamefont {Olsen}\ and\ \citenamefont
  {Bradley}(2008)}]{Olsen2008}%
  \BibitemOpen
  \bibfield  {author} {\bibinfo {author} {\bibfnamefont {M.~K.}\ \bibnamefont
  {Olsen}}\ and\ \bibinfo {author} {\bibfnamefont {A.~S.}\ \bibnamefont
  {Bradley}},\ }\href {\doibase 10.1103/PhysRevA.77.023813} {\bibfield
  {journal} {\bibinfo  {journal} {Phys. Rev. A}\ }\textbf {\bibinfo {volume}
  {77}},\ \bibinfo {pages} {023813} (\bibinfo {year} {2008})}\BibitemShut
  {NoStop}%
\bibitem [{\citenamefont {Midgley}\ \emph {et~al.}(2010)\citenamefont
  {Midgley}, \citenamefont {Ferris},\ and\ \citenamefont
  {Olsen}}]{Midgley2010}%
  \BibitemOpen
  \bibfield  {author} {\bibinfo {author} {\bibfnamefont {S.~L.~W.}\
  \bibnamefont {Midgley}}, \bibinfo {author} {\bibfnamefont {A.~J.}\
  \bibnamefont {Ferris}}, \ and\ \bibinfo {author} {\bibfnamefont {M.~K.}\
  \bibnamefont {Olsen}},\ }\href {\doibase 10.1103/PhysRevA.81.022101}
  {\bibfield  {journal} {\bibinfo  {journal} {Phys. Rev. A}\ }\textbf {\bibinfo
  {volume} {81}},\ \bibinfo {pages} {022101} (\bibinfo {year}
  {2010})}\BibitemShut {NoStop}%
\bibitem [{\citenamefont {Olsen}(2013)}]{Olsen2013}%
  \BibitemOpen
  \bibfield  {author} {\bibinfo {author} {\bibfnamefont {M.~K.}\ \bibnamefont
  {Olsen}},\ }\href {\doibase 10.1103/PhysRevA.88.051802} {\bibfield  {journal}
  {\bibinfo  {journal} {Phys. Rev. A}\ }\textbf {\bibinfo {volume} {88}},\
  \bibinfo {pages} {051802(R)} (\bibinfo {year} {2013})}\BibitemShut {NoStop}%
\bibitem [{\citenamefont {Schneeloch}\ \emph {et~al.}(2013)\citenamefont
  {Schneeloch}, \citenamefont {Broadbent}, \citenamefont {Walborn},
  \citenamefont {Cavalcanti},\ and\ \citenamefont {Howell}}]{Schneeloch2013}%
  \BibitemOpen
  \bibfield  {author} {\bibinfo {author} {\bibfnamefont {J.}~\bibnamefont
  {Schneeloch}}, \bibinfo {author} {\bibfnamefont {C.~J.}\ \bibnamefont
  {Broadbent}}, \bibinfo {author} {\bibfnamefont {S.~P.}\ \bibnamefont
  {Walborn}}, \bibinfo {author} {\bibfnamefont {E.~G.}\ \bibnamefont
  {Cavalcanti}}, \ and\ \bibinfo {author} {\bibfnamefont {J.~C.}\ \bibnamefont
  {Howell}},\ }\href {\doibase 10.1103/PhysRevA.87.062103} {\bibfield
  {journal} {\bibinfo  {journal} {Phys. Rev. A}\ }\textbf {\bibinfo {volume}
  {87}},\ \bibinfo {pages} {062103} (\bibinfo {year} {2013})}\BibitemShut
  {NoStop}%
\bibitem [{\citenamefont {He}\ and\ \citenamefont {Reid}(2013)}]{He2013}%
  \BibitemOpen
  \bibfield  {author} {\bibinfo {author} {\bibfnamefont {Q.~Y.}\ \bibnamefont
  {He}}\ and\ \bibinfo {author} {\bibfnamefont {M.~D.}\ \bibnamefont {Reid}},\
  }\href {\doibase 10.1103/PhysRevA.88.052121} {\bibfield  {journal} {\bibinfo
  {journal} {Phys. Rev. A}\ }\textbf {\bibinfo {volume} {88}},\ \bibinfo
  {pages} {052121} (\bibinfo {year} {2013})}\BibitemShut {NoStop}%
\bibitem [{\citenamefont {He}\ and\ \citenamefont {Ficek}(2014)}]{He2014}%
  \BibitemOpen
  \bibfield  {author} {\bibinfo {author} {\bibfnamefont {Q.~Y.}\ \bibnamefont
  {He}}\ and\ \bibinfo {author} {\bibfnamefont {Z.}~\bibnamefont {Ficek}},\
  }\href {\doibase 10.1103/PhysRevA.89.022332} {\bibfield  {journal} {\bibinfo
  {journal} {Phys. Rev. A}\ }\textbf {\bibinfo {volume} {89}},\ \bibinfo
  {pages} {022332} (\bibinfo {year} {2014})}\BibitemShut {NoStop}%
\bibitem [{\citenamefont {Evans}\ and\ \citenamefont
  {Wiseman}(2014)}]{Evans2014}%
  \BibitemOpen
  \bibfield  {author} {\bibinfo {author} {\bibfnamefont {D.~A.}\ \bibnamefont
  {Evans}}\ and\ \bibinfo {author} {\bibfnamefont {H.~M.}\ \bibnamefont
  {Wiseman}},\ }\href {\doibase 10.1103/PhysRevA.90.012114} {\bibfield
  {journal} {\bibinfo  {journal} {Phys. Rev. A}\ }\textbf {\bibinfo {volume}
  {90}},\ \bibinfo {pages} {012114} (\bibinfo {year} {2014})}\BibitemShut
  {NoStop}%
\bibitem [{\citenamefont {Bowles}\ \emph {et~al.}(2014)\citenamefont {Bowles},
  \citenamefont {V{\'{e}}rtesi}, \citenamefont {Quintino},\ and\ \citenamefont
  {Brunner}}]{Bowles2014}%
  \BibitemOpen
  \bibfield  {author} {\bibinfo {author} {\bibfnamefont {J.}~\bibnamefont
  {Bowles}}, \bibinfo {author} {\bibfnamefont {T.}~\bibnamefont
  {V{\'{e}}rtesi}}, \bibinfo {author} {\bibfnamefont {M.~T.}\ \bibnamefont
  {Quintino}}, \ and\ \bibinfo {author} {\bibfnamefont {N.}~\bibnamefont
  {Brunner}},\ }\href {\doibase 10.1103/PhysRevLett.112.200402} {\bibfield
  {journal} {\bibinfo  {journal} {Phys. Rev. Lett.}\ }\textbf {\bibinfo
  {volume} {112}},\ \bibinfo {pages} {200402} (\bibinfo {year}
  {2014})}\BibitemShut {NoStop}%
\bibitem [{\citenamefont {Skrzypczyk}\ \emph {et~al.}(2014)\citenamefont
  {Skrzypczyk}, \citenamefont {Navascu{\'{e}}s},\ and\ \citenamefont
  {Cavalcanti}}]{Skrzypczyk2014}%
  \BibitemOpen
  \bibfield  {author} {\bibinfo {author} {\bibfnamefont {P.}~\bibnamefont
  {Skrzypczyk}}, \bibinfo {author} {\bibfnamefont {M.}~\bibnamefont
  {Navascu{\'{e}}s}}, \ and\ \bibinfo {author} {\bibfnamefont {D.}~\bibnamefont
  {Cavalcanti}},\ }\href {\doibase 10.1103/PhysRevLett.112.180404} {\bibfield
  {journal} {\bibinfo  {journal} {Phys. Rev. Lett.}\ }\textbf {\bibinfo
  {volume} {112}},\ \bibinfo {pages} {180404} (\bibinfo {year}
  {2014})}\BibitemShut {NoStop}%
\bibitem [{\citenamefont {Wang}\ \emph {et~al.}(2014)\citenamefont {Wang},
  \citenamefont {Gong}, \citenamefont {Ficek},\ and\ \citenamefont
  {He}}]{Wang2014}%
  \BibitemOpen
  \bibfield  {author} {\bibinfo {author} {\bibfnamefont {M.}~\bibnamefont
  {Wang}}, \bibinfo {author} {\bibfnamefont {Q.~H.}\ \bibnamefont {Gong}},
  \bibinfo {author} {\bibfnamefont {Z.}~\bibnamefont {Ficek}}, \ and\ \bibinfo
  {author} {\bibfnamefont {Q.~Y.}\ \bibnamefont {He}},\ }\href {\doibase
  10.1103/PhysRevA.90.023801} {\bibfield  {journal} {\bibinfo  {journal} {Phys.
  Rev. A}\ }\textbf {\bibinfo {volume} {90}},\ \bibinfo {pages} {023801}
  (\bibinfo {year} {2014})}\BibitemShut {NoStop}%
\bibitem [{\citenamefont {Rosales-Z\'{a}rate}\ \emph
  {et~al.}(2015)\citenamefont {Rosales-Z\'{a}rate}, \citenamefont {Teh},
  \citenamefont {Kiesewetter}, \citenamefont {Brolis}, \citenamefont {Ng},\
  and\ \citenamefont {Reid}}]{Reidjosab2015}%
  \BibitemOpen
  \bibfield  {author} {\bibinfo {author} {\bibfnamefont {L.}~\bibnamefont
  {Rosales-Z\'{a}rate}}, \bibinfo {author} {\bibfnamefont {R.~Y.}\ \bibnamefont
  {Teh}}, \bibinfo {author} {\bibfnamefont {S.}~\bibnamefont {Kiesewetter}},
  \bibinfo {author} {\bibfnamefont {A.}~\bibnamefont {Brolis}}, \bibinfo
  {author} {\bibfnamefont {K.}~\bibnamefont {Ng}}, \ and\ \bibinfo {author}
  {\bibfnamefont {M.~D.}\ \bibnamefont {Reid}},\ }\href {\doibase
  10.1364/JOSAB.32.000A82} {\bibfield  {journal} {\bibinfo  {journal} {J. Opt.
  Soc. Am. B}\ }\textbf {\bibinfo {volume} {32}},\ \bibinfo {pages} {A82}
  (\bibinfo {year} {2015})}\BibitemShut {NoStop}%
\bibitem [{\citenamefont {He}\ \emph {et~al.}(2015{\natexlab{b}})\citenamefont
  {He}, \citenamefont {Gong},\ and\ \citenamefont {Reid}}]{QHe2015}%
  \BibitemOpen
  \bibfield  {author} {\bibinfo {author} {\bibfnamefont {Q.~Y.}\ \bibnamefont
  {He}}, \bibinfo {author} {\bibfnamefont {Q.~H.}\ \bibnamefont {Gong}}, \ and\
  \bibinfo {author} {\bibfnamefont {M.~D.}\ \bibnamefont {Reid}},\ }\href
  {\doibase 10.1103/PhysRevLett.114.060402} {\bibfield  {journal} {\bibinfo
  {journal} {Phys. Rev. Lett.}\ }\textbf {\bibinfo {volume} {114}},\ \bibinfo
  {pages} {060402} (\bibinfo {year} {2015}{\natexlab{b}})}\BibitemShut
  {NoStop}%
\bibitem [{\citenamefont {Tan}\ \emph {et~al.}(2015)\citenamefont {Tan},
  \citenamefont {Zhang},\ and\ \citenamefont {Li}}]{Tan2015}%
  \BibitemOpen
  \bibfield  {author} {\bibinfo {author} {\bibfnamefont {H.~T.}\ \bibnamefont
  {Tan}}, \bibinfo {author} {\bibfnamefont {X.~C.}\ \bibnamefont {Zhang}}, \
  and\ \bibinfo {author} {\bibfnamefont {G.~X.}\ \bibnamefont {Li}},\ }\href
  {\doibase 10.1103/PhysRevA.91.032121} {\bibfield  {journal} {\bibinfo
  {journal} {Phys. Rev. A}\ }\textbf {\bibinfo {volume} {91}},\ \bibinfo
  {pages} {032121} (\bibinfo {year} {2015})}\BibitemShut {NoStop}%
\bibitem [{\citenamefont {Quintino}\ \emph {et~al.}(2015)\citenamefont
  {Quintino}, \citenamefont {V\'ertesi}, \citenamefont {Cavalcanti},
  \citenamefont {Augusiak}, \citenamefont {Demianowicz}, \citenamefont
  {Ac\'{i}n},\ and\ \citenamefont {Brunner}}]{Marco2015}%
  \BibitemOpen
  \bibfield  {author} {\bibinfo {author} {\bibfnamefont {M.~T.}\ \bibnamefont
  {Quintino}}, \bibinfo {author} {\bibfnamefont {T.}~\bibnamefont {V\'ertesi}},
  \bibinfo {author} {\bibfnamefont {D.}~\bibnamefont {Cavalcanti}}, \bibinfo
  {author} {\bibfnamefont {R.}~\bibnamefont {Augusiak}}, \bibinfo {author}
  {\bibfnamefont {M.}~\bibnamefont {Demianowicz}}, \bibinfo {author}
  {\bibfnamefont {A.}~\bibnamefont {Ac\'{i}n}}, \ and\ \bibinfo {author}
  {\bibfnamefont {N.}~\bibnamefont {Brunner}},\ }\href {\doibase
  10.1103/PhysRevA.92.032107} {\bibfield  {journal} {\bibinfo  {journal} {Phys.
  Rev. A}\ }\textbf {\bibinfo {volume} {92}},\ \bibinfo {pages} {032107}
  (\bibinfo {year} {2015})}\BibitemShut {NoStop}%
\bibitem [{\citenamefont {Bowles}\ \emph {et~al.}(2016)\citenamefont {Bowles},
  \citenamefont {Hirsch}, \citenamefont {Quintino},\ and\ \citenamefont
  {Brunner}}]{Bowles2016}%
  \BibitemOpen
  \bibfield  {author} {\bibinfo {author} {\bibfnamefont {J.}~\bibnamefont
  {Bowles}}, \bibinfo {author} {\bibfnamefont {F.}~\bibnamefont {Hirsch}},
  \bibinfo {author} {\bibfnamefont {M.~T.}\ \bibnamefont {Quintino}}, \ and\
  \bibinfo {author} {\bibfnamefont {N.}~\bibnamefont {Brunner}},\ }\href
  {\doibase 10.1103/PhysRevA.93.022121} {\bibfield  {journal} {\bibinfo
  {journal} {Phys. Rev. A}\ }\textbf {\bibinfo {volume} {93}},\ \bibinfo
  {pages} {022121} (\bibinfo {year} {2016})}\BibitemShut {NoStop}%
\bibitem [{\citenamefont {Olsen}(2017)}]{Olsen2017}%
  \BibitemOpen
  \bibfield  {author} {\bibinfo {author} {\bibfnamefont {M.~K.}\ \bibnamefont
  {Olsen}},\ }\href {\doibase 10.1103/PhysRevLett.119.160501} {\bibfield
  {journal} {\bibinfo  {journal} {Phys. Rev. Lett.}\ }\textbf {\bibinfo
  {volume} {119}},\ \bibinfo {pages} {160501} (\bibinfo {year}
  {2017})}\BibitemShut {NoStop}%
\bibitem [{\citenamefont {Baker}\ \emph {et~al.}(2018)\citenamefont {Baker},
  \citenamefont {Wollmann},\ and\ \citenamefont {Pryde}}]{Baker2018}%
  \BibitemOpen
  \bibfield  {author} {\bibinfo {author} {\bibfnamefont {T.~J.}\ \bibnamefont
  {Baker}}, \bibinfo {author} {\bibfnamefont {S.}~\bibnamefont {Wollmann}}, \
  and\ \bibinfo {author} {\bibfnamefont {G.~J.}\ \bibnamefont {Pryde}},\ }\href
  {\doibase 10.1088/2040-8986/aaaa3c} {\bibfield  {journal} {\bibinfo
  {journal} {J. Opt.}\ }\textbf {\bibinfo {volume} {20}},\ \bibinfo {pages}
  {034008} (\bibinfo {year} {2018})}\BibitemShut {NoStop}%
\bibitem [{\citenamefont {Qin}\ \emph {et~al.}(2017)\citenamefont {Qin},
  \citenamefont {Deng}, \citenamefont {Tian}, \citenamefont {Wang},
  \citenamefont {Su}, \citenamefont {Xie},\ and\ \citenamefont
  {Peng}}]{Qin2017}%
  \BibitemOpen
  \bibfield  {author} {\bibinfo {author} {\bibfnamefont {Z.~Z.}\ \bibnamefont
  {Qin}}, \bibinfo {author} {\bibfnamefont {X.~W.}\ \bibnamefont {Deng}},
  \bibinfo {author} {\bibfnamefont {C.~X.}\ \bibnamefont {Tian}}, \bibinfo
  {author} {\bibfnamefont {M.~H.}\ \bibnamefont {Wang}}, \bibinfo {author}
  {\bibfnamefont {X.~L.}\ \bibnamefont {Su}}, \bibinfo {author} {\bibfnamefont
  {C.~D.}\ \bibnamefont {Xie}}, \ and\ \bibinfo {author} {\bibfnamefont
  {K.~C.}\ \bibnamefont {Peng}},\ }\href {\doibase 10.1103/PhysRevA.95.052114}
  {\bibfield  {journal} {\bibinfo  {journal} {Phys. Rev. A}\ }\textbf {\bibinfo
  {volume} {95}},\ \bibinfo {pages} {052114} (\bibinfo {year}
  {2017})}\BibitemShut {NoStop}%
\bibitem [{\citenamefont {Sun}\ \emph {et~al.}(2016)\citenamefont {Sun},
  \citenamefont {Ye}, \citenamefont {Xu}, \citenamefont {Xu}, \citenamefont
  {Tang}, \citenamefont {Wu}, \citenamefont {Chen}, \citenamefont {Li},\ and\
  \citenamefont {Guo}}]{Sun2016}%
  \BibitemOpen
  \bibfield  {author} {\bibinfo {author} {\bibfnamefont {K.}~\bibnamefont
  {Sun}}, \bibinfo {author} {\bibfnamefont {X.~J.}\ \bibnamefont {Ye}},
  \bibinfo {author} {\bibfnamefont {J.~S.}\ \bibnamefont {Xu}}, \bibinfo
  {author} {\bibfnamefont {X.~Y.}\ \bibnamefont {Xu}}, \bibinfo {author}
  {\bibfnamefont {J.~S.}\ \bibnamefont {Tang}}, \bibinfo {author}
  {\bibfnamefont {Y.~C.}\ \bibnamefont {Wu}}, \bibinfo {author} {\bibfnamefont
  {J.~L.}\ \bibnamefont {Chen}}, \bibinfo {author} {\bibfnamefont {C.~F.}\
  \bibnamefont {Li}}, \ and\ \bibinfo {author} {\bibfnamefont {G.~C.}\
  \bibnamefont {Guo}},\ }\href {\doibase 10.1103/PhysRevLett.116.160404}
  {\bibfield  {journal} {\bibinfo  {journal} {Phys. Rev. Lett.}\ }\textbf
  {\bibinfo {volume} {116}},\ \bibinfo {pages} {160404} (\bibinfo {year}
  {2016})}\BibitemShut {NoStop}%
\bibitem [{\citenamefont {Wollmann}\ \emph {et~al.}(2016)\citenamefont
  {Wollmann}, \citenamefont {Walk}, \citenamefont {Bennet}, \citenamefont
  {Wiseman},\ and\ \citenamefont {Pryde}}]{Wollmann2016}%
  \BibitemOpen
  \bibfield  {author} {\bibinfo {author} {\bibfnamefont {S.}~\bibnamefont
  {Wollmann}}, \bibinfo {author} {\bibfnamefont {N.}~\bibnamefont {Walk}},
  \bibinfo {author} {\bibfnamefont {A.~J.}\ \bibnamefont {Bennet}}, \bibinfo
  {author} {\bibfnamefont {H.~M.}\ \bibnamefont {Wiseman}}, \ and\ \bibinfo
  {author} {\bibfnamefont {G.~J.}\ \bibnamefont {Pryde}},\ }\href {\doibase
  10.1103/PhysRevLett.116.160403} {\bibfield  {journal} {\bibinfo  {journal}
  {Phys. Rev. Lett.}\ }\textbf {\bibinfo {volume} {116}},\ \bibinfo {pages}
  {160403} (\bibinfo {year} {2016})}\BibitemShut {NoStop}%
\bibitem [{\citenamefont {Xiao}\ \emph {et~al.}(2017)\citenamefont {Xiao},
  \citenamefont {Ye}, \citenamefont {Sun}, \citenamefont {Xu}, \citenamefont
  {Li},\ and\ \citenamefont {Guo}}]{Xiao2017}%
  \BibitemOpen
  \bibfield  {author} {\bibinfo {author} {\bibfnamefont {Y.}~\bibnamefont
  {Xiao}}, \bibinfo {author} {\bibfnamefont {X.~J.}\ \bibnamefont {Ye}},
  \bibinfo {author} {\bibfnamefont {K.}~\bibnamefont {Sun}}, \bibinfo {author}
  {\bibfnamefont {J.~S.}\ \bibnamefont {Xu}}, \bibinfo {author} {\bibfnamefont
  {C.~F.}\ \bibnamefont {Li}}, \ and\ \bibinfo {author} {\bibfnamefont {G.~C.}\
  \bibnamefont {Guo}},\ }\href {\doibase 10.1103/PhysRevLett.118.140404}
  {\bibfield  {journal} {\bibinfo  {journal} {Phys. Rev. Lett.}\ }\textbf
  {\bibinfo {volume} {118}},\ \bibinfo {pages} {140404} (\bibinfo {year}
  {2017})}\BibitemShut {NoStop}%
\bibitem [{\citenamefont {Tischler}\ \emph {et~al.}(2018)\citenamefont
  {Tischler}, \citenamefont {Ghafari}, \citenamefont {Baker}, \citenamefont
  {Slussarenko}, \citenamefont {Patel}, \citenamefont {Weston}, \citenamefont
  {Wollmann}, \citenamefont {Shalm}, \citenamefont {Verma}, \citenamefont
  {Nam}, \citenamefont {Nguyen}, \citenamefont {Wiseman},\ and\ \citenamefont
  {Pryde}}]{Tischler2018}%
  \BibitemOpen
  \bibfield  {author} {\bibinfo {author} {\bibfnamefont {N.}~\bibnamefont
  {Tischler}}, \bibinfo {author} {\bibfnamefont {F.}~\bibnamefont {Ghafari}},
  \bibinfo {author} {\bibfnamefont {T.~J.}\ \bibnamefont {Baker}}, \bibinfo
  {author} {\bibfnamefont {S.}~\bibnamefont {Slussarenko}}, \bibinfo {author}
  {\bibfnamefont {R.~B.}\ \bibnamefont {Patel}}, \bibinfo {author}
  {\bibfnamefont {M.~M.}\ \bibnamefont {Weston}}, \bibinfo {author}
  {\bibfnamefont {S.}~\bibnamefont {Wollmann}}, \bibinfo {author}
  {\bibfnamefont {L.~K.}\ \bibnamefont {Shalm}}, \bibinfo {author}
  {\bibfnamefont {V.~B.}\ \bibnamefont {Verma}}, \bibinfo {author}
  {\bibfnamefont {S.~W.}\ \bibnamefont {Nam}}, \bibinfo {author} {\bibfnamefont
  {H.~C.}\ \bibnamefont {Nguyen}}, \bibinfo {author} {\bibfnamefont {H.~M.}\
  \bibnamefont {Wiseman}}, \ and\ \bibinfo {author} {\bibfnamefont {G.~J.}\
  \bibnamefont {Pryde}},\ }\href {\doibase 10.1103/PhysRevLett.121.100401}
  {\bibfield  {journal} {\bibinfo  {journal} {Phys. Rev. Lett.}\ }\textbf
  {\bibinfo {volume} {121}},\ \bibinfo {pages} {100401} (\bibinfo {year}
  {2018})}\BibitemShut {NoStop}%
\bibitem [{\citenamefont {Fadel}\ \emph {et~al.}(2018)\citenamefont {Fadel},
  \citenamefont {Zibold}, \citenamefont {D\'{e}camps},\ and\ \citenamefont
  {Treutlein}}]{Fadel2018}%
  \BibitemOpen
  \bibfield  {author} {\bibinfo {author} {\bibfnamefont {M.}~\bibnamefont
  {Fadel}}, \bibinfo {author} {\bibfnamefont {T.}~\bibnamefont {Zibold}},
  \bibinfo {author} {\bibfnamefont {B.}~\bibnamefont {D\'{e}camps}}, \ and\
  \bibinfo {author} {\bibfnamefont {P.}~\bibnamefont {Treutlein}},\ }\href
  {\doibase 10.1126/science.aao1850} {\bibfield  {journal} {\bibinfo  {journal}
  {Science}\ }\textbf {\bibinfo {volume} {360}},\ \bibinfo {pages} {409}
  (\bibinfo {year} {2018})}\BibitemShut {NoStop}%
\bibitem [{\citenamefont {Restrepo}\ \emph {et~al.}(2017)\citenamefont
  {Restrepo}, \citenamefont {Favero},\ and\ \citenamefont
  {Ciuti}}]{Restrepo2017}%
  \BibitemOpen
  \bibfield  {author} {\bibinfo {author} {\bibfnamefont {J.}~\bibnamefont
  {Restrepo}}, \bibinfo {author} {\bibfnamefont {I.}~\bibnamefont {Favero}}, \
  and\ \bibinfo {author} {\bibfnamefont {C.}~\bibnamefont {Ciuti}},\ }\href
  {\doibase 10.1103/PhysRevA.95.023832} {\bibfield  {journal} {\bibinfo
  {journal} {Phys. Rev. A}\ }\textbf {\bibinfo {volume} {95}},\ \bibinfo
  {pages} {023832} (\bibinfo {year} {2017})}\BibitemShut {NoStop}%
\bibitem [{\citenamefont {Sun}\ \emph {et~al.}(2017)\citenamefont {Sun},
  \citenamefont {Mao}, \citenamefont {Dai}, \citenamefont {Ficek},
  \citenamefont {He},\ and\ \citenamefont {Gong}}]{Sun2017}%
  \BibitemOpen
  \bibfield  {author} {\bibinfo {author} {\bibfnamefont {F.~X.}\ \bibnamefont
  {Sun}}, \bibinfo {author} {\bibfnamefont {D.}~\bibnamefont {Mao}}, \bibinfo
  {author} {\bibfnamefont {Y.~T.}\ \bibnamefont {Dai}}, \bibinfo {author}
  {\bibfnamefont {Z.}~\bibnamefont {Ficek}}, \bibinfo {author} {\bibfnamefont
  {Q.~Y.}\ \bibnamefont {He}}, \ and\ \bibinfo {author} {\bibfnamefont {Q.~H.}\
  \bibnamefont {Gong}},\ }\href {\doibase 10.1088/1367-2630/aa9c9a} {\bibfield
  {journal} {\bibinfo  {journal} {New J. Phy.}\ }\textbf {\bibinfo {volume}
  {19}},\ \bibinfo {pages} {123039} (\bibinfo {year} {2017})}\BibitemShut
  {NoStop}%
\bibitem [{\citenamefont {Wang}\ \emph {et~al.}(2017)\citenamefont {Wang},
  \citenamefont {Liu}, \citenamefont {Wu},\ and\ \citenamefont
  {Liu}}]{Wang2017}%
  \BibitemOpen
  \bibfield  {author} {\bibinfo {author} {\bibfnamefont {C.}~\bibnamefont
  {Wang}}, \bibinfo {author} {\bibfnamefont {Y.~L.}\ \bibnamefont {Liu}},
  \bibinfo {author} {\bibfnamefont {R.}~\bibnamefont {Wu}}, \ and\ \bibinfo
  {author} {\bibfnamefont {Y.~X.}\ \bibnamefont {Liu}},\ }\href {\doibase
  10.1103/PhysRevA.96.013818} {\bibfield  {journal} {\bibinfo  {journal} {Phys.
  Rev. A}\ }\textbf {\bibinfo {volume} {96}},\ \bibinfo {pages} {013818}
  (\bibinfo {year} {2017})}\BibitemShut {NoStop}%
\bibitem [{\citenamefont {{Moaddel Haghighi}}\ \emph
  {et~al.}(2018)\citenamefont {{Moaddel Haghighi}}, \citenamefont {Malossi},
  \citenamefont {Natali}, \citenamefont {{Di Giuseppe}},\ and\ \citenamefont
  {Vitali}}]{MoaddelHaghighi2018}%
  \BibitemOpen
  \bibfield  {author} {\bibinfo {author} {\bibfnamefont {I.}~\bibnamefont
  {{Moaddel Haghighi}}}, \bibinfo {author} {\bibfnamefont {N.}~\bibnamefont
  {Malossi}}, \bibinfo {author} {\bibfnamefont {R.}~\bibnamefont {Natali}},
  \bibinfo {author} {\bibfnamefont {G.}~\bibnamefont {{Di Giuseppe}}}, \ and\
  \bibinfo {author} {\bibfnamefont {D.}~\bibnamefont {Vitali}},\ }\href
  {\doibase 10.1103/PhysRevApplied.9.034031} {\bibfield  {journal} {\bibinfo
  {journal} {Phys. Rev. Applied}\ }\textbf {\bibinfo {volume} {9}},\ \bibinfo
  {pages} {034031} (\bibinfo {year} {2018})}\BibitemShut {NoStop}%
\bibitem [{\citenamefont {Metelmann}\ and\ \citenamefont
  {Clerk}(2015)}]{Metelmann2015}%
  \BibitemOpen
  \bibfield  {author} {\bibinfo {author} {\bibfnamefont {A.}~\bibnamefont
  {Metelmann}}\ and\ \bibinfo {author} {\bibfnamefont {A.~A.}\ \bibnamefont
  {Clerk}},\ }\href {\doibase 10.1103/PhysRevX.5.021025} {\bibfield  {journal}
  {\bibinfo  {journal} {Phys. Rev. X}\ }\textbf {\bibinfo {volume} {5}},\
  \bibinfo {pages} {021025} (\bibinfo {year} {2015})}\BibitemShut {NoStop}%
\bibitem [{\citenamefont {Xu}\ and\ \citenamefont {Li}(2015)}]{Xu2015}%
  \BibitemOpen
  \bibfield  {author} {\bibinfo {author} {\bibfnamefont {X.~W.}\ \bibnamefont
  {Xu}}\ and\ \bibinfo {author} {\bibfnamefont {Y.}~\bibnamefont {Li}},\ }\href
  {\doibase 10.1103/PhysRevA.91.053854} {\bibfield  {journal} {\bibinfo
  {journal} {Phys. Rev. A}\ }\textbf {\bibinfo {volume} {91}},\ \bibinfo
  {pages} {053854} (\bibinfo {year} {2015})}\BibitemShut {NoStop}%
\bibitem [{\citenamefont {Xu}\ \emph {et~al.}(2016)\citenamefont {Xu},
  \citenamefont {Li}, \citenamefont {Chen},\ and\ \citenamefont
  {Liu}}]{Xu2016}%
  \BibitemOpen
  \bibfield  {author} {\bibinfo {author} {\bibfnamefont {X.~W.}\ \bibnamefont
  {Xu}}, \bibinfo {author} {\bibfnamefont {Y.}~\bibnamefont {Li}}, \bibinfo
  {author} {\bibfnamefont {A.~X.}\ \bibnamefont {Chen}}, \ and\ \bibinfo
  {author} {\bibfnamefont {Y.~X.}\ \bibnamefont {Liu}},\ }\href {\doibase
  10.1103/PhysRevA.93.023827} {\bibfield  {journal} {\bibinfo  {journal} {Phys.
  Rev. A}\ }\textbf {\bibinfo {volume} {93}},\ \bibinfo {pages} {023827}
  (\bibinfo {year} {2016})}\BibitemShut {NoStop}%
\bibitem [{\citenamefont {Xu}\ \emph {et~al.}(2017)\citenamefont {Xu},
  \citenamefont {Chen}, \citenamefont {Li},\ and\ \citenamefont
  {Liu}}]{Xu2017}%
  \BibitemOpen
  \bibfield  {author} {\bibinfo {author} {\bibfnamefont {X.~W.}\ \bibnamefont
  {Xu}}, \bibinfo {author} {\bibfnamefont {A.~X.}\ \bibnamefont {Chen}},
  \bibinfo {author} {\bibfnamefont {Y.}~\bibnamefont {Li}}, \ and\ \bibinfo
  {author} {\bibfnamefont {Y.~X.}\ \bibnamefont {Liu}},\ }\href {\doibase
  10.1103/PhysRevA.96.053853} {\bibfield  {journal} {\bibinfo  {journal} {Phys.
  Rev. A}\ }\textbf {\bibinfo {volume} {96}},\ \bibinfo {pages} {053853}
  (\bibinfo {year} {2017})}\BibitemShut {NoStop}%
\bibitem [{\citenamefont {Ruesink}\ \emph {et~al.}(2016)\citenamefont
  {Ruesink}, \citenamefont {Miri}, \citenamefont {Al{\`{u}}},\ and\
  \citenamefont {Verhagen}}]{Ruesink2016}%
  \BibitemOpen
  \bibfield  {author} {\bibinfo {author} {\bibfnamefont {F.}~\bibnamefont
  {Ruesink}}, \bibinfo {author} {\bibfnamefont {M.-A.}\ \bibnamefont {Miri}},
  \bibinfo {author} {\bibfnamefont {A.}~\bibnamefont {Al{\`{u}}}}, \ and\
  \bibinfo {author} {\bibfnamefont {E.}~\bibnamefont {Verhagen}},\ }\href
  {\doibase 10.1038/ncomms13662} {\bibfield  {journal} {\bibinfo  {journal}
  {Nat. Commun.}\ }\textbf {\bibinfo {volume} {7}},\ \bibinfo {pages} {13662}
  (\bibinfo {year} {2016})}\BibitemShut {NoStop}%
\bibitem [{\citenamefont {Barzanjeh}\ \emph {et~al.}(2017)\citenamefont
  {Barzanjeh}, \citenamefont {Wulf}, \citenamefont {Peruzzo}, \citenamefont
  {Kalaee}, \citenamefont {Dieterle}, \citenamefont {Painter},\ and\
  \citenamefont {Fink}}]{Barzanjeh2017}%
  \BibitemOpen
  \bibfield  {author} {\bibinfo {author} {\bibfnamefont {S.}~\bibnamefont
  {Barzanjeh}}, \bibinfo {author} {\bibfnamefont {M.}~\bibnamefont {Wulf}},
  \bibinfo {author} {\bibfnamefont {M.}~\bibnamefont {Peruzzo}}, \bibinfo
  {author} {\bibfnamefont {M.}~\bibnamefont {Kalaee}}, \bibinfo {author}
  {\bibfnamefont {P.~B.}\ \bibnamefont {Dieterle}}, \bibinfo {author}
  {\bibfnamefont {O.}~\bibnamefont {Painter}}, \ and\ \bibinfo {author}
  {\bibfnamefont {J.~M.}\ \bibnamefont {Fink}},\ }\href {\doibase
  10.1038/s41467-017-01304-x} {\bibfield  {journal} {\bibinfo  {journal} {Nat.
  Commun.}\ }\textbf {\bibinfo {volume} {8}},\ \bibinfo {pages} {953} (\bibinfo
  {year} {2017})}\BibitemShut {NoStop}%
\bibitem [{\citenamefont {Bernier}\ \emph {et~al.}(2017)\citenamefont
  {Bernier}, \citenamefont {T{\'{o}}th}, \citenamefont {Koottandavida},
  \citenamefont {Ioannou}, \citenamefont {Malz}, \citenamefont {Nunnenkamp},
  \citenamefont {Feofanov},\ and\ \citenamefont {Kippenberg}}]{Bernier2017}%
  \BibitemOpen
  \bibfield  {author} {\bibinfo {author} {\bibfnamefont {N.~R.}\ \bibnamefont
  {Bernier}}, \bibinfo {author} {\bibfnamefont {L.~D.}\ \bibnamefont
  {T{\'{o}}th}}, \bibinfo {author} {\bibfnamefont {A.}~\bibnamefont
  {Koottandavida}}, \bibinfo {author} {\bibfnamefont {M.~A.}\ \bibnamefont
  {Ioannou}}, \bibinfo {author} {\bibfnamefont {D.}~\bibnamefont {Malz}},
  \bibinfo {author} {\bibfnamefont {A.}~\bibnamefont {Nunnenkamp}}, \bibinfo
  {author} {\bibfnamefont {A.~K.}\ \bibnamefont {Feofanov}}, \ and\ \bibinfo
  {author} {\bibfnamefont {T.~J.}\ \bibnamefont {Kippenberg}},\ }\href
  {\doibase 10.1038/s41467-017-00447-1} {\bibfield  {journal} {\bibinfo
  {journal} {Nat. Commun.}\ }\textbf {\bibinfo {volume} {8}},\ \bibinfo {pages}
  {604} (\bibinfo {year} {2017})}\BibitemShut {NoStop}%
\bibitem [{\citenamefont {Yuan}\ and\ \citenamefont {Wang}(2017)}]{Yuan2017}%
  \BibitemOpen
  \bibfield  {author} {\bibinfo {author} {\bibfnamefont {H.~Y.}\ \bibnamefont
  {Yuan}}\ and\ \bibinfo {author} {\bibfnamefont {X.~R.}\ \bibnamefont
  {Wang}},\ }\href {\doibase 10.1063/1.4977083} {\bibfield  {journal} {\bibinfo
   {journal} {Appl. Phys. Lett.}\ }\textbf {\bibinfo {volume} {110}},\ \bibinfo
  {pages} {082403} (\bibinfo {year} {2017})}\BibitemShut {NoStop}%
\bibitem [{\citenamefont {Dejesus}\ and\ \citenamefont
  {Kaufman}(1987)}]{Dejesus1987}%
  \BibitemOpen
  \bibfield  {author} {\bibinfo {author} {\bibfnamefont {E.~X.}\ \bibnamefont
  {Dejesus}}\ and\ \bibinfo {author} {\bibfnamefont {C.}~\bibnamefont
  {Kaufman}},\ }\href {\doibase 10.1103/PhysRevA.35.5288} {\bibfield  {journal}
  {\bibinfo  {journal} {Phys. Rev. A}\ }\textbf {\bibinfo {volume} {35}},\
  \bibinfo {pages} {5288} (\bibinfo {year} {1987})}\BibitemShut {NoStop}%
\bibitem [{\citenamefont {Genes}\ \emph {et~al.}(2008)\citenamefont {Genes},
  \citenamefont {Mari}, \citenamefont {Tombesi},\ and\ \citenamefont
  {Vitali}}]{Genes2008}%
  \BibitemOpen
  \bibfield  {author} {\bibinfo {author} {\bibfnamefont {C.}~\bibnamefont
  {Genes}}, \bibinfo {author} {\bibfnamefont {A.}~\bibnamefont {Mari}},
  \bibinfo {author} {\bibfnamefont {P.}~\bibnamefont {Tombesi}}, \ and\
  \bibinfo {author} {\bibfnamefont {D.}~\bibnamefont {Vitali}},\ }\href
  {\doibase 10.1103/PhysRevA.78.032316} {\bibfield  {journal} {\bibinfo
  {journal} {Phys. Rev. A}\ }\textbf {\bibinfo {volume} {78}},\ \bibinfo
  {pages} {032316} (\bibinfo {year} {2008})}\BibitemShut {NoStop}%
\bibitem [{\citenamefont {Barzanjeh}\ \emph {et~al.}(2011)\citenamefont
  {Barzanjeh}, \citenamefont {Vitali}, \citenamefont {Tombesi},\ and\
  \citenamefont {Milburn}}]{Barzanjeh2011}%
  \BibitemOpen
  \bibfield  {author} {\bibinfo {author} {\bibfnamefont {S.}~\bibnamefont
  {Barzanjeh}}, \bibinfo {author} {\bibfnamefont {D.}~\bibnamefont {Vitali}},
  \bibinfo {author} {\bibfnamefont {P.}~\bibnamefont {Tombesi}}, \ and\
  \bibinfo {author} {\bibfnamefont {G.~J.}\ \bibnamefont {Milburn}},\ }\href
  {\doibase 10.1103/PhysRevA.84.042342} {\bibfield  {journal} {\bibinfo
  {journal} {Phys. Rev. A}\ }\textbf {\bibinfo {volume} {84}},\ \bibinfo
  {pages} {042342} (\bibinfo {year} {2011})}\BibitemShut {NoStop}%
\bibitem [{\citenamefont {Vidal}\ and\ \citenamefont
  {Werner}(2002)}]{Vidal2002}%
  \BibitemOpen
  \bibfield  {author} {\bibinfo {author} {\bibfnamefont {G.}~\bibnamefont
  {Vidal}}\ and\ \bibinfo {author} {\bibfnamefont {R.~F.}\ \bibnamefont
  {Werner}},\ }\href {\doibase 10.1103/PhysRevA.65.032314} {\bibfield
  {journal} {\bibinfo  {journal} {Phys. Rev. A}\ }\textbf {\bibinfo {volume}
  {65}},\ \bibinfo {pages} {032314} (\bibinfo {year} {2002})}\BibitemShut
  {NoStop}%
\bibitem [{\citenamefont {Adesso}\ \emph {et~al.}(2004)\citenamefont {Adesso},
  \citenamefont {Serafini},\ and\ \citenamefont {Illuminati}}]{Adesso2004}%
  \BibitemOpen
  \bibfield  {author} {\bibinfo {author} {\bibfnamefont {G.}~\bibnamefont
  {Adesso}}, \bibinfo {author} {\bibfnamefont {A.}~\bibnamefont {Serafini}}, \
  and\ \bibinfo {author} {\bibfnamefont {F.}~\bibnamefont {Illuminati}},\
  }\href {\doibase 10.1103/PhysRevA.70.022318} {\bibfield  {journal} {\bibinfo
  {journal} {Phys. Rev. A}\ }\textbf {\bibinfo {volume} {70}},\ \bibinfo
  {pages} {022318} (\bibinfo {year} {2004})}\BibitemShut {NoStop}%
\bibitem [{\citenamefont {Kogias}\ \emph {et~al.}(2015)\citenamefont {Kogias},
  \citenamefont {Lee}, \citenamefont {Ragy},\ and\ \citenamefont
  {Adesso}}]{Kogias2015}%
  \BibitemOpen
  \bibfield  {author} {\bibinfo {author} {\bibfnamefont {I.}~\bibnamefont
  {Kogias}}, \bibinfo {author} {\bibfnamefont {A.~R.}\ \bibnamefont {Lee}},
  \bibinfo {author} {\bibfnamefont {S.}~\bibnamefont {Ragy}}, \ and\ \bibinfo
  {author} {\bibfnamefont {G.}~\bibnamefont {Adesso}},\ }\href {\doibase
  10.1103/PhysRevLett.114.060403} {\bibfield  {journal} {\bibinfo  {journal}
  {Phys. Rev. Lett.}\ }\textbf {\bibinfo {volume} {114}},\ \bibinfo {pages}
  {060403} (\bibinfo {year} {2015})}\BibitemShut {NoStop}%
\bibitem [{\citenamefont {Hillery}\ and\ \citenamefont
  {Zubairy}(2006)}]{Hillery2006}%
  \BibitemOpen
  \bibfield  {author} {\bibinfo {author} {\bibfnamefont {M.}~\bibnamefont
  {Hillery}}\ and\ \bibinfo {author} {\bibfnamefont {M.~S.}\ \bibnamefont
  {Zubairy}},\ }\href {\doibase 10.1103/PhysRevLett.96.050503} {\bibfield
  {journal} {\bibinfo  {journal} {Phys. Rev. Lett.}\ }\textbf {\bibinfo
  {volume} {96}},\ \bibinfo {pages} {050503} (\bibinfo {year}
  {2006})}\BibitemShut {NoStop}%
\bibitem [{\citenamefont {Cavalcanti}\ \emph {et~al.}(2011)\citenamefont
  {Cavalcanti}, \citenamefont {He}, \citenamefont {Reid},\ and\ \citenamefont
  {Wiseman}}]{Cavalcanti2011}%
  \BibitemOpen
  \bibfield  {author} {\bibinfo {author} {\bibfnamefont {E.~G.}\ \bibnamefont
  {Cavalcanti}}, \bibinfo {author} {\bibfnamefont {Q.~Y.}\ \bibnamefont {He}},
  \bibinfo {author} {\bibfnamefont {M.~D.}\ \bibnamefont {Reid}}, \ and\
  \bibinfo {author} {\bibfnamefont {H.~M.}\ \bibnamefont {Wiseman}},\ }\href
  {\doibase 10.1103/PhysRevA.84.032115} {\bibfield  {journal} {\bibinfo
  {journal} {Phys. Rev. A}\ }\textbf {\bibinfo {volume} {84}},\ \bibinfo
  {pages} {032115} (\bibinfo {year} {2011})}\BibitemShut {NoStop}%
\end{thebibliography}%


%
\end{document}